\documentclass[10pt]{iopart}

\usepackage{amssymb}
\usepackage{pstricks}
\usepackage{epsfig}
\usepackage{color}

\def\<{\left<}
\def\>{\right>}
\def\ket|#1>{\left|#1\right>}
\def\bra<#1|{\left<#1\right|}
\def\elem<#1|#2|#3>{\left<#1\right|#2\left|#3\right>}
\def\({\left(}
\def\){\right)}

\def\Z{{\mathbb Z}}
\def\C{{\mathbb C}}
\def\R{{\mathbb R}}

\def\rp(#1,#2)#3{\rput(#1,#2){\small #3}}
\newgray{llgray}{.9}

\begin{document}

\title{Qubism: self-similar visualization of many-body
  wavefunctions}

\author{Javier Rodr\'{\i}guez-Laguna$^{1,2}$, Piotr Migda\l{$^1$},
  Miguel Ib\'a\~nez Berganza$^3$, Maciej Lewenstein$^{1,4}$ and Germ\'an
  Sierra $^3$} 

\address{$^1$ ICFO--Institute of Photonic Sciences, Castelldefels
  (Barcelona), Spain}

\address{$^2$ Mathematics Department, Universidad Carlos III de
  Madrid, Spain}

\address{$^3$ IFT--Instituto de F\'{\i}sica Te\'orica, UAM/CSIC,
  Cantoblanco (Madrid), Spain}

\address{$^4$ ICREA–-Instituci\'o Catalana de Recerca i Estudis Avan\c cats,
  Barcelona, Spain}
 
\ead{javier.rodriguez@icfo.es}

\date{December 13, 2011}

\begin{abstract}
A visualization scheme for quantum many-body wavefunctions is
described, which we have termed {\em qubism}. Its main property is its
{\em recursivity}: increasing the number of qubits reflects in an
increase in the image resolution. Thus, the plots are typically
fractal. As examples, we provide images for the ground states of
commonly used Hamiltonians in condensed matter and cold atom physics,
such as Heisenberg or ITF. Many features of the wavefunction, such as
magnetization, correlations and criticality, can be visualized as
properties of the images. In particular, factorizability can be easily
spotted, and a way to estimate the entanglement entropy from the image
is provided.
\end{abstract}




\section{\label{introd}Introduction}

\subsection{Motivation}

Most of the difficulty of quantum many-body physics stems from the
complexity of its fundamental mathematical objects: many-body
wavefunctions and density matrices. In the simplest case, where we
have $N$ qubits, a wavefunction (pure state) can be considered as a
function mapping $\{0,1\}^N \mapsto \C$. Therefore, it is
characterized by $2^N$ complex parameters. Density matrices (mixed
states) have even greater mathematical complexity, mapping $\{0,1\}^N
\times \{0,1\}^N \mapsto \C$, i.e. $2^{2N}$ complex parameters. 

The aim of this work is to describe a pictorial representation of
quantum many-body wavefunctions, in which a wavefunction
characterizing a chain of $N$ qubits maps into an image with
$2^{N/2}\times 2^{N/2}$ pixels. Thus, an increase in the number of
qubits reflects itself in an increase in the resolution of the
image. These images are typically fractal, and sometimes
self-similar. Extension to higher spin {\em qudits} is
straightforward, and is also explored. Some physical properties of the
wavefunction become visually apprehensible: magnetization (ferro or
antiferromagnetic character), criticality, entanglement, translation
invariance, permutation invariance, etc.

\subsection{Historical review}

Visualization of complex data is a common problem in many branches of
science and technology. Let us review here some of the relevant
hallmarks that preceded our work.

Historically, it can be argued that the single most relevant advance
in calculus was the discovery of the relation between algebraic
functions and curves in the plane in the \textsc{xvii}
century. Function visualization provided an insight which guided most
of the subsequent development of calculus, not only by helping solve
established problems, but also by suggesting new ones. With the advent
of the new information technologies, complex data visualization has
developed into a full-fledged field of research. The reader is
directed to \cite{Handbook} for a recent review of state-of-the-art
techniques, and \cite{Tufte} for a historical perspective.

As a relevant example, the problem of visualization of DNA and protein
sequences was addressed in 1990 by Jeffrey making use of the so-called
{\em chaos game representation} (CGR) \cite{jeffrey_90}. DNA sequences
are long, highly correlated strings of four symbols,
$\{A,C,G,T\}$. Let us label the four corners of a square with
them. Now, select the central point of the square and proceed as
follows. Pick the next symbol from the string. Find the point midway
between the selected point and the corner which corresponds to the
symbol. Mark that point, and make it your new selected point. If the
sequence is genuinely random, the points will cover the square
uniformly. Otherwise, patterns will emerge, very often with fractal
structure. The original purpose of the technique was mere
visualization, but it evolved \cite{almeida_01} to provide
quantitative measurements, such as Shannon entropies, which help
researchers to characterize DNA and protein sequences \cite{liu_07}.

In 2000, Hao and coworkers \cite{hao_00} developed a different
representation technique for long DNA sequences that also had fractal
properties. Given a certain value of $N$, they computed the frequency
of every subsequence of length $N$ within the global sequence, thus
obtaining a mathematical object which is similar to a many-body
wavefunction, only mapping from $\{A,C,G,T\}^N \mapsto \R$. The number
of different subsequences of length $N$ is $4^N$. Hao and coworkers
represented the subsequence probability distribution by dividing a
unit square in a recursive way, into $4^N$ small squares, and
attaching a color to each of them. The resulting images have fractal
appearance, as remarked by the authors, but their quantification is
not pursued. Their purpose is to identify which types of subsequences
are under-represented, and to this end they analyse the corresponding
patterns of low frequency.

In 2005 Latorre \cite{latorre_05}, unaware of the work of Hao et al.,
developed independently a mapping between bitmap images and many-body
wavefunctions which has a similar philosophy, and applied quantum
information techniques in order to develop an image compression
algorithm. Although the compression rate was not competitive with
standard {\em jpeg}, the insight provided by the mapping was of high
value \cite{Le_11}. A crucial insight for the present work was the
idea that Latorre's mapping might be inverted, in order to obtain
bitmap images out of many-body wavefunctions.

Focusing on quantum mechanics, the simplest visualization technique is
provided by the representation of a qubit as a point on a Bloch
sphere. Early work of Ettore Majorana \cite{Majorana_32} proved that a
permutation-invariant system of $N$ spins-1/2 can be represented as a
set of $n$ points on the Bloch sphere. This Majorana representation
has proved very useful in characterizations of entanglement
\cite{Aulbach_10,Ganczarek_12}.

A different approach that can provide visualization schemes of quantum
many-body systems was introduced by Wootters and coworkers in 2004
\cite{Wootters.04}. The idea is to set a bidimensional array of
operators which fulfill certain properties, and measure their
expectation values in the given state. Those values, displayed in a 2D
lattice, generate a discrete analogue of a Wigner function.

\subsection{Plan of this work}

In this work, we describe a set of techniques which provide graphical
representations of many-body wavefunctions, which share many features
with the schemes of Latorre and Hao and coworkers. The main insight is
that the increase in complexity as we add more qubits is mapped into
an increase in the resolution of the corresponding image. Thus, the
thermodynamic limit, when the number of qubits tends to infinity,
corresponds to the continuum limit for the images. The scheme is
recursive in scales, and this makes the images look fractal in a
natural way. In fact, as we will discuss, exact self-similarity of the
image implies that the wavefunction is factorizable.

In section \ref{2dplots} we describe the basic wavefunction plotting
scheme, while section \ref{examples} is devoted to providing several
examples (Heisenberg, ITF, Dicke states, product states, etc.)
emphasizing how physical features map into plot features.  The
procedure is generalized in section \ref{otherplots}, and some
alternative plotting schemes are described, which allow us to try
states of spin-1 systems, such as the AKLT state. Section
\ref{selfsim}, on the other hand, deals with the fractal properties of
the plots and extracts useful information from them. In section
\ref{entanglement} discusses how to recognize entangled states in a
wavefunction plot, along with a simple technique to estimate
entanglement by inspection. A different plotting scheme, based upon
the frame representation and related to the Wootters group approach is
succintly described in section \ref{pauli}, and a few pictures are
provided for the sake of comparison. The article finishes with
conclusions and a description of future work.


\section{\label{2dplots}2D-plot of many-body wavefunctions}

\begin{figure}
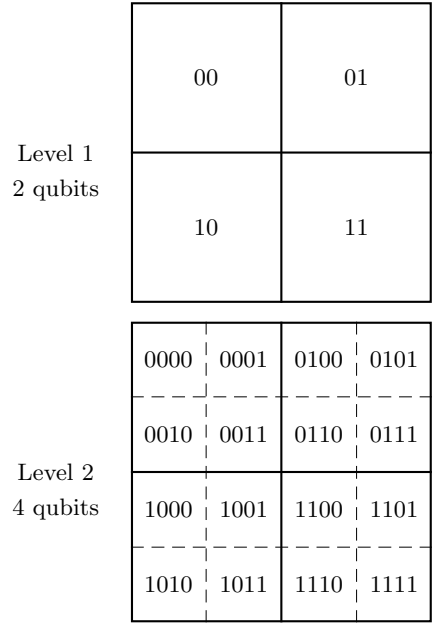

\psset{unit=1mm}
\rput(45,0){%
\rp(-10,-20){Level 1}
\rp(-10,-25){2 qubits}
\psframe(0,0)(40,-40)\psline(0,-20)(40,-20)\psline(20,0)(20,-40)%
\rp(10,-10){00}
\rp(10,-30){10}
\rp(30,-10){01}
\rp(30,-30){11}
}
\rput(45,-42.5){%
\rp(-10,-20){Level 2}
\rp(-10,-25){4 qubits}
\psframe(0,0)(40,-40)
\psline(0,-20)(40,-20)\psline(20,0)(20,-40)%
\psline[linestyle=dashed,linewidth=.2pt](10,0)(10,-40)%
\psline[linestyle=dashed,linewidth=.2pt](30,0)(30,-40)%
\psline[linestyle=dashed,linewidth=.2pt](0,-10)(40,-10)%
\psline[linestyle=dashed,linewidth=.2pt](0,-30)(40,-30)%
\rput(5,-5){\rp(0,0){0000}
\rp(10,0){0001}
\rp(20,0){0100}
\rp(30,0){0101}}
\rput(5,-15){\rp(0,0){0010}
\rp(10,0){0011}
\rp(20,0){0110}
\rp(30,0){0111}}
\rput(5,-25){\rp(0,0){1000}
\rp(10,0){1001}
\rp(20,0){1100}
\rp(30,0){1101}}
\rput(5,-35){\rp(0,0){1010}
\rp(10,0){1011}
\rp(20,0){1110}
\rp(30,0){1111}}
}
\vspace{85mm}
\caption{\label{fig.2dplot} 2D plotting scheme of many-body
  wavefunctions. Top: each of the tensor basis states for 2 qubits:
  $\ket|00>$, $\ket|01>$, $\ket|10>$ and $\ket|11>$ is mapped into one
  of the four level-1 squares. Bottom: mapping of 4-qubit basis states
  into level-2 squares.}
\end{figure}

Let us consider a couple of qubits. The tensor basis is composed of
four states: $\ket|00>$, $\ket|01>$, $\ket|10>$ and
$\ket|11>$. Consider also a unit square, $[0,1]\times[0,1]$, and divide it into
four ``level-1'' squares. We can associate each of the basis states to
one of the squares, as shown in figure \ref{fig.2dplot} (top).

The basic mapping is, therefore:

\begin{equation}
\begin{array}{cc}
00 \to \hbox{Upper left} & 01 \to  \hbox{Upper right} \\
10 \to \hbox{Lower left} & 11 \to  \hbox{Lower right} \\
\end{array}
\label{mapping1}
\end{equation}

The splitting of squares can be iterated, obtaining level-2 squares,
etc., as it is shown in figure \ref{fig.2dplot} (bottom). For a
wavefunction with $N$ qubits, we will have to descend down to
level-$N/2$ squares. Each of them will correspond to one of the tensor
basis states. If the number of qubits $N$ is odd, the same scheme can
be applied with a rectangular plot. The last step is straightforward:
attach a color, or a gray level, to each of the level-$N/2$ squares,
depending on the value of the wavefunction. Obviously, using only
levels of gray (or color intensity), only real values can be attached
easily to each tensor basis state. In order to show phases, we
recourse to a color-cycle scheme.

\begin{figure}
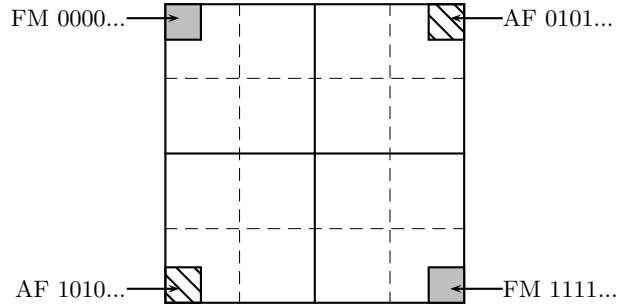

\psset{unit=1mm}
\rput(40,-3){%
\psframe(0,0)(40,-40)\psline(0,-20)(40,-20)\psline(20,0)(20,-40)%
\psline[linestyle=dashed,linewidth=.2pt](10,0)(10,-40)%
\psline[linestyle=dashed,linewidth=.2pt](30,0)(30,-40)%
\psline[linestyle=dashed,linewidth=.2pt](0,-10)(40,-10)%
\psline[linestyle=dashed,linewidth=.2pt](0,-30)(40,-30)%
\psframe[fillstyle=solid,fillcolor=lightgray](0,0)(5,-5)
\psframe[fillstyle=solid,fillcolor=lightgray](40,-40)(35,-35)
\psline{<-}(2,-2)(-5,-2)\rput[r](-5,-2){\small FM 0000...}
\psline{<-}(38,-38)(45,-38)\rput[l](45,-38){\small FM 1111...}
\psframe[fillstyle=vlines,fillcolor=lightgray](35,0)(40,-5)
\psframe[fillstyle=vlines,fillcolor=lightgray](0,-35)(5,-40)
\psline{<-}(2,-38)(-5,-38)\rput[r](-5,-38){\small AF 1010...}
\psline{<-}(38,-2)(45,-2)\rput[l](45,-2){\small AF 0101...}
}
\vspace{45mm}
\caption{\label{fig.2dplot.features} Exemplification of some features
  of the 2D-plot scheme. The FM states $0000...$ and $1111...$
  correspond to the NW and SE corners of the image, respectively. The
  NE and SW corners, on the other hand, correspond to the N\'eel AF
  states. The $\Z_2$ symmetry operation $0\leftrightarrow 1$
  corresponds to a rotation by $180^\circ$.}
\end{figure}

Figure \ref{fig.2dplot.features} shows some features of this
mapping. The ferromagnetic (FM) states, $0000...$ and $1111...$
correspond, respectively, to the upper-left (NW) and lower-right (SE)
corners of the image, while the N\'eel antiferromagnetic (AF) states
correspond to the other two corners: $0101...$ is the upper-right (NE)
corner, and $1010...$ is the lower-left one (SW). It is
straightforward to realize that the $\Z_2$ symmetry operation
$0\leftrightarrow 1$ corresponds to a rotation of $180^\circ$ around
the center of the plot.

Let us consider any state $s\in\{0,1\}^N$ and denote its bits by
$s=\{X_1 Y_1 X_2 Y_2 \cdots X_n Y_n\}$, with $n=N/2$. In order to find
the point in the unit square where this state will be mapped, build
the following numbers:

\begin{equation}
x=\sum_{i=1}^n X_i 2^{-i}, \qquad y=\sum_{i=1}^n Y_i 2^{-i}
\label{coord.2d}
\end{equation}

\noindent Those are the coordinates of the upper-left corner of the
corresponding level-$n$ square, if $(0,0)$ is the upper-left corner of
the square, and the $y$-coordinate grows downwards.

In our plots, unless otherwise stated, each cell is filled with a
color corresponding with its wave-function amplitude according to the
following scheme: color intensity corresponds to the modulus (white
means zero), and hue is used as a phase indicator. Concretely, red is
used for positive values and green for negative ones, with a smooth
interpolation scheme.

\begin{figure}
\begin{center}
\begin{tabular}{ccccc}
\epsfig{file=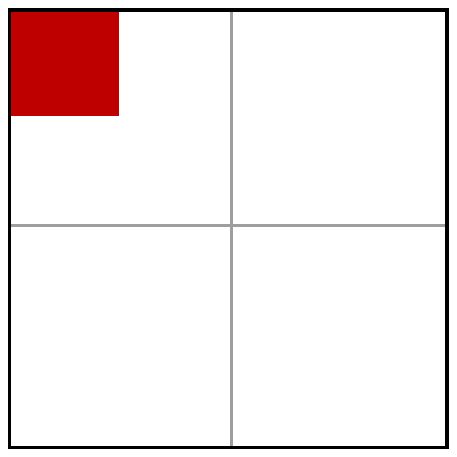,width=1.5cm} &
\epsfig{file=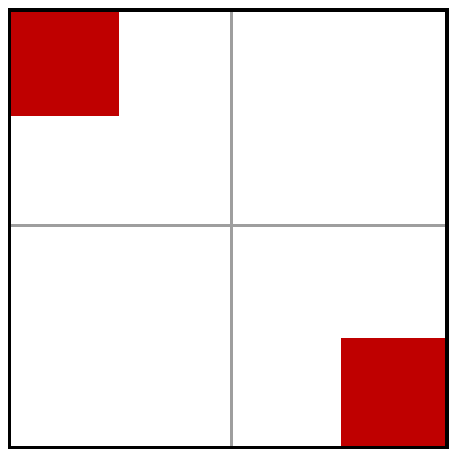,width=1.5cm} &
\epsfig{file=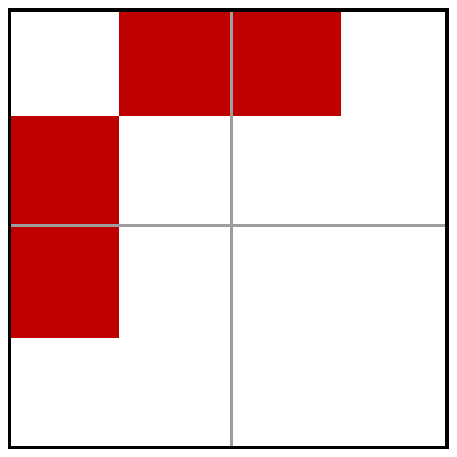,width=1.5cm} &
\epsfig{file=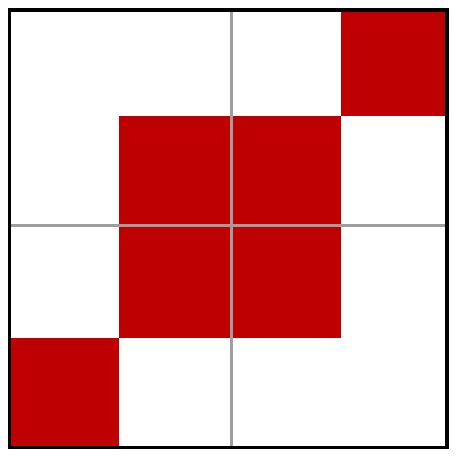,width=1.5cm} \\
(A) &
(B) &
(C) &
(D) \\
\end{tabular}
\end{center}
\caption{\label{fig.n4} Qubistic plots of some important states with
  $N=4$ qubits. (A) The simplest factorizable state $\ket|0000>$. (B)
  The GHZ state. (C) W state. (D) Dicke state at half-filling.}
\end{figure}

Figure \ref{fig.n4} provides some simple examples of states with $N=4$
qubits. Panel (A) is the qubistic plot for the factorizable state
$\ket|0000>$, in which only the upper-left corner cell of the plot is
marked. In (B) the Greenberger–Horne–Zeilinger (GHZ) state is shown,
$\ket|0000>+\ket|1111>$. In this case, two opposite corner cells are
marked, with the same color since their relative phase is
positive. The third panel, fig. \ref{fig.n4} (C) corresponds to the
so-called W state for $N=4$ qubits, i.e.:
$\ket|1000>+\ket|0100>+\ket|0010>+\ket|0001>$. The plot consists of
$N$ marked cells distributed along the upper and leftmost rows of the
plot. For larger values of $N$ the spacing among these marked cells
becomes exponential. The fourth plot in figure \ref{fig.n4} is the
Dicke state at half-filling, i.e.: the linear combination, with equal
weights, of all basis states with half the qubits $1$. 

\section{\label{examples}Examples of Qubistic 2D-plots}

Along this section we will study qubistic plots of low-energy states
of hamiltonians which are relevant in condensed matter physics and
ultracold atomic cases, giving special attention to quantum phase
transitions (QPT) \cite{Sachdev,Lewenstein}.

\subsection{Heisenberg Ground State: Spin Liquid structure}

Our next example will be taken from the low-energy spectrum of the
antiferromagnetic (AF) spin-$1/2$ Heisenberg model in 1D with periodic
boundary conditions (PBC).

\begin{equation}
H= \sum_{i=1}^N {\vec S}_i\cdot {\vec S}_{i+1}
\label{heis.model}
\end{equation}

\begin{figure}
\begin{center}
\begin{tabular}{ccc}
\multicolumn{3}{c}{\epsfig{file=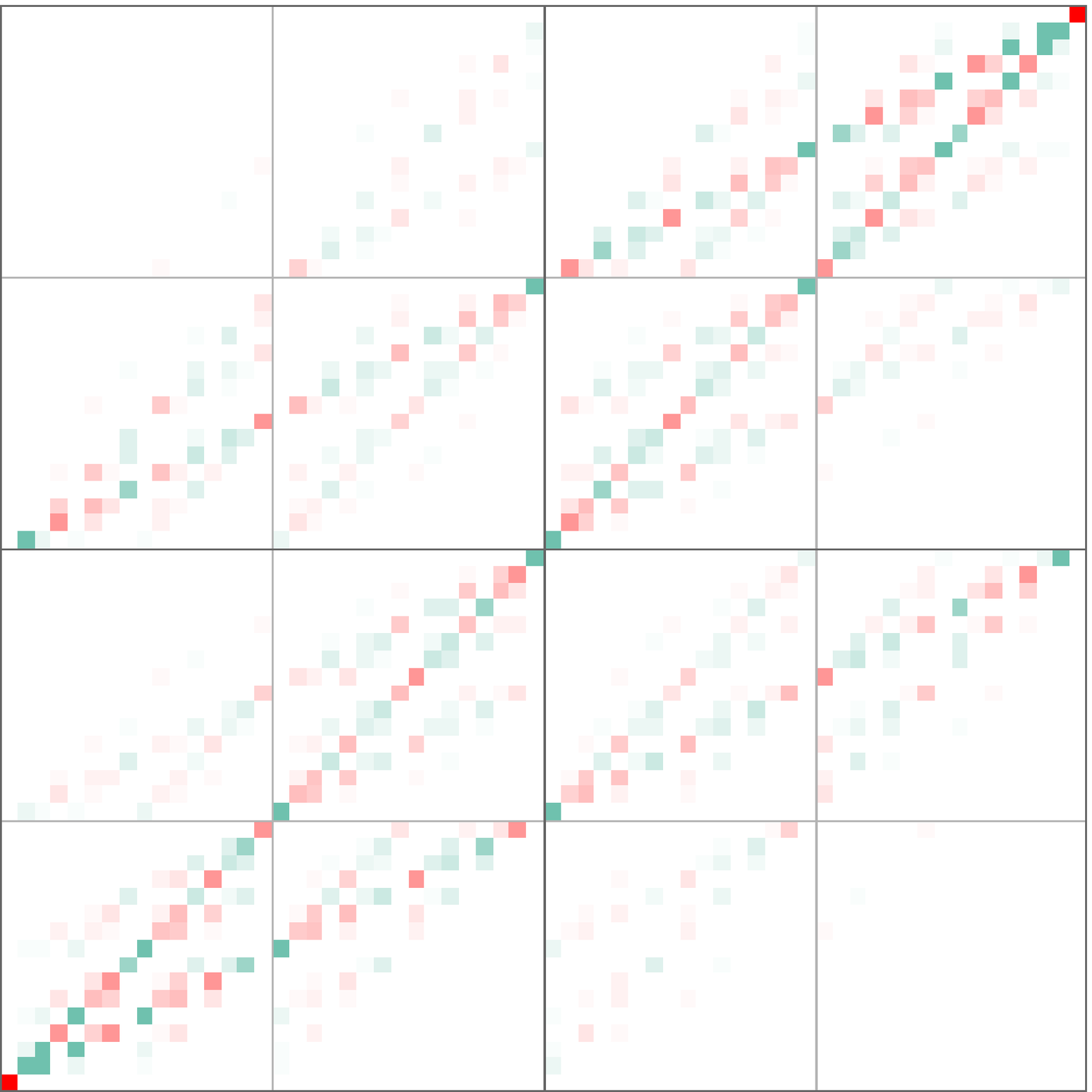,width=6cm}} \\
\epsfig{file=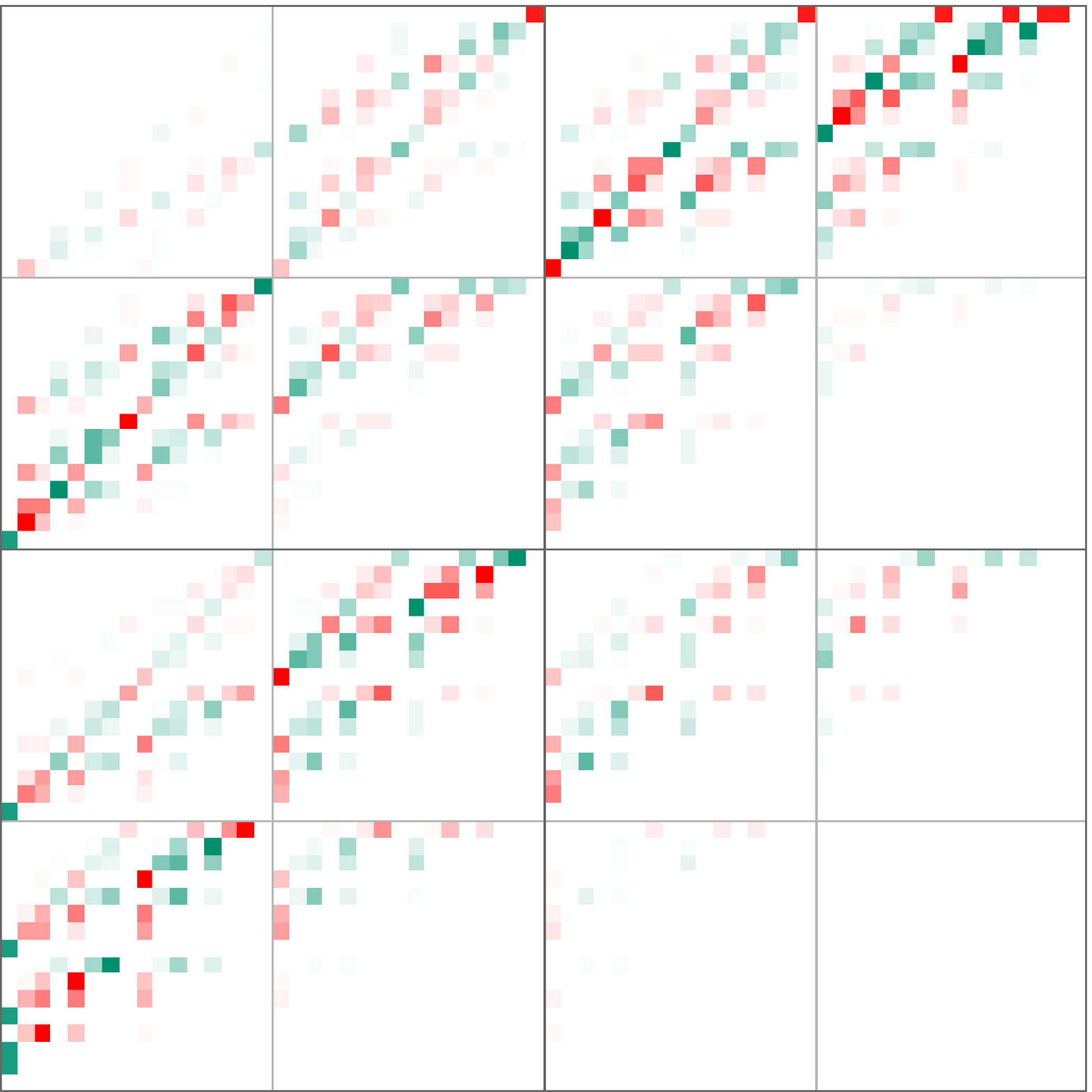,width=2.7cm} &
\epsfig{file=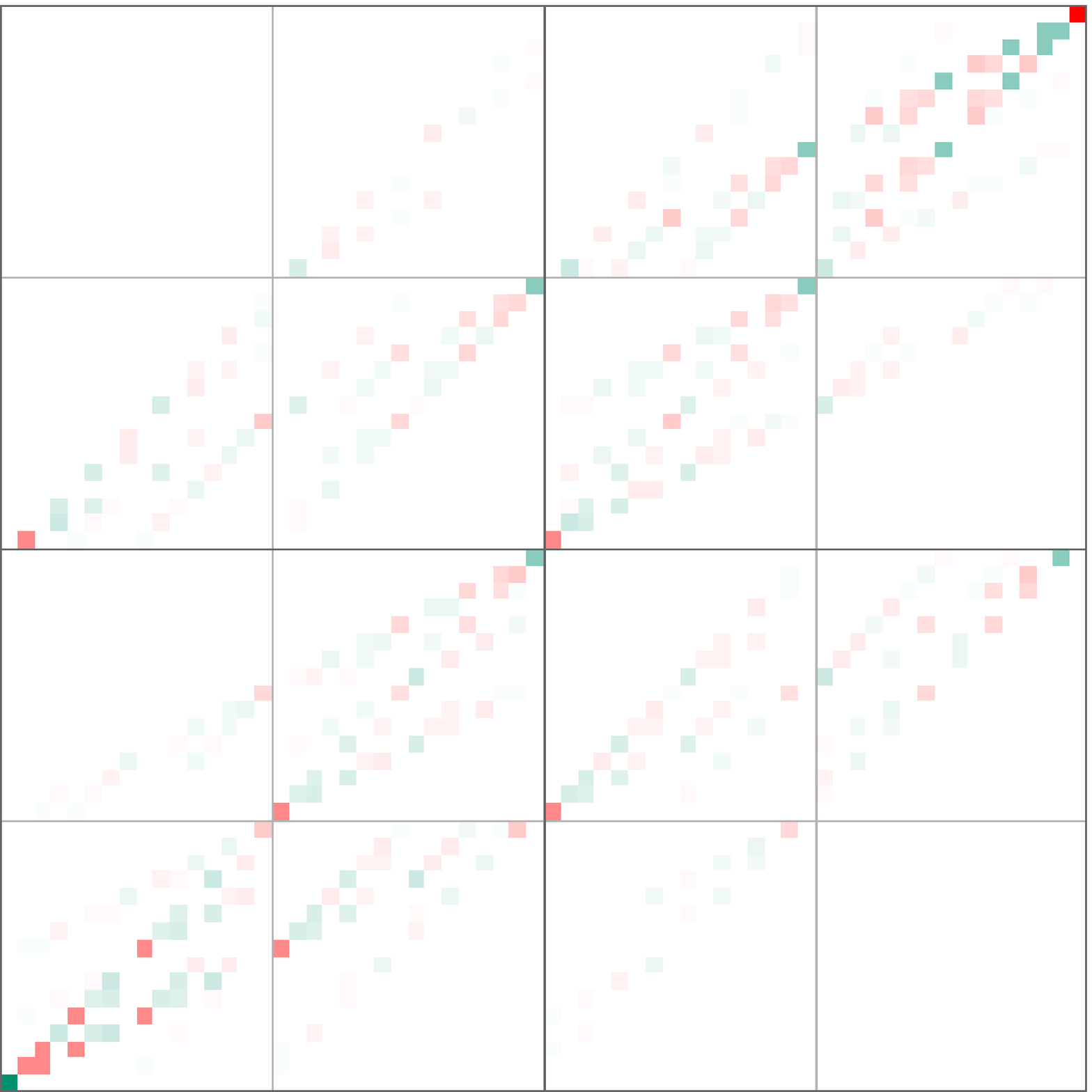,width=2.7cm} &
\epsfig{file=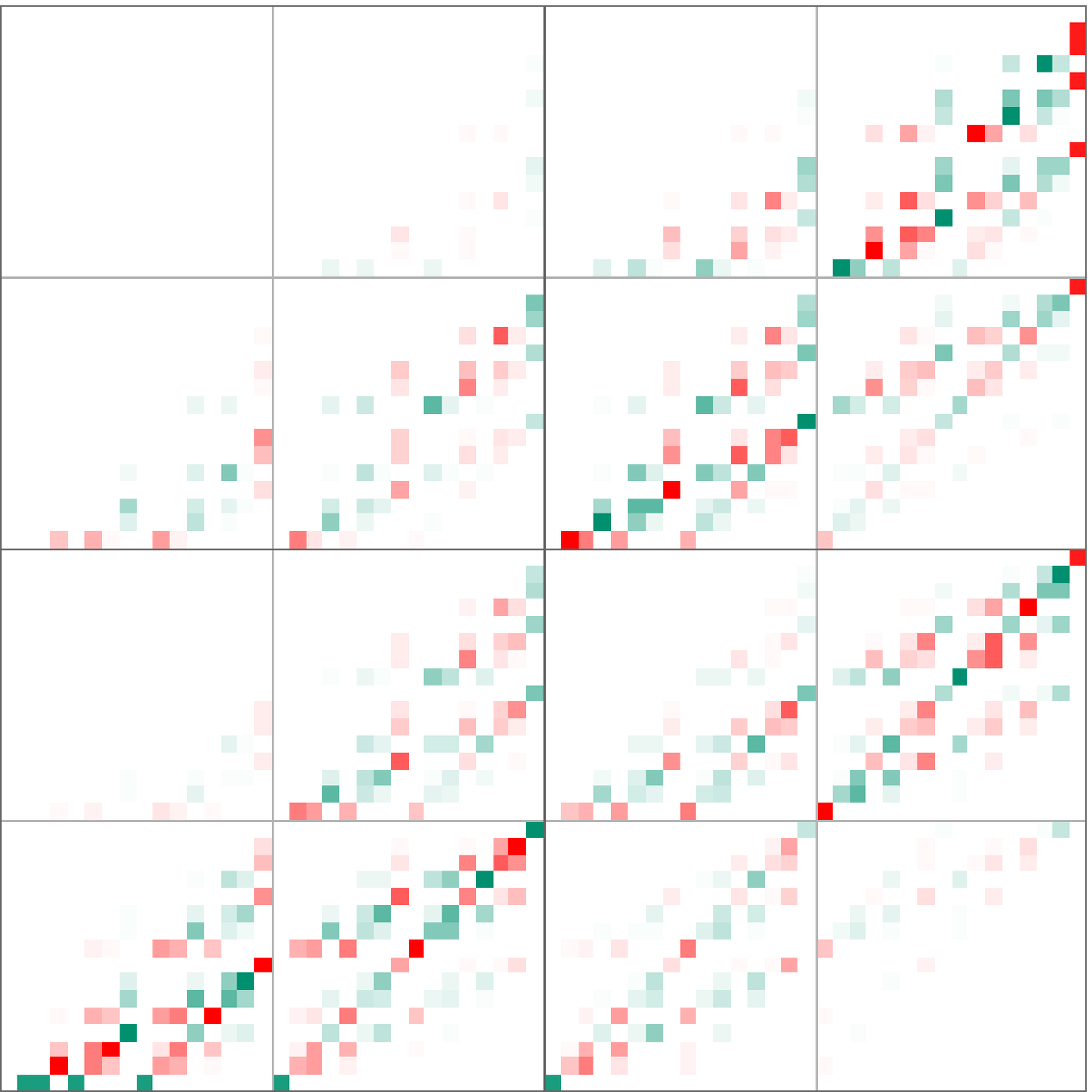,width=2.7cm} \\
\end{tabular}
\end{center}
\caption{\label{fig.heis} Top: Ground state of the 1D spin-$1/2$ AF
  Heisenberg model with PBC for 12 spins; Bottom: the lowest energy
  excitations, which make up a triplet. White means zero probability,
  color intensity reflects the modulus of the wavefunction amplitude,
  while color hue marks the phase.}
\end{figure}

The top panel of figure \ref{fig.heis} shows the ground state of
equation \ref{heis.model}, while the bottom row shows the first three
excited states, which constitute a spin-1 triplet.

Let us focus on the ground state (figure \ref{fig.heis}, top
panel). The most salient feature is its intense diagonal line, joining
the two N\'eel states, which get maximal weight. The states conforming
that diagonal are all made up of pairs $01$ and $10$, in any
order. This main diagonal is the depiction of a set of {\em pairwise
  singlet bonds}: $(1,2)(3,4)\cdots(N-1,N)$.

\begin{figure}{c}
\begin{center}
\epsfig{file=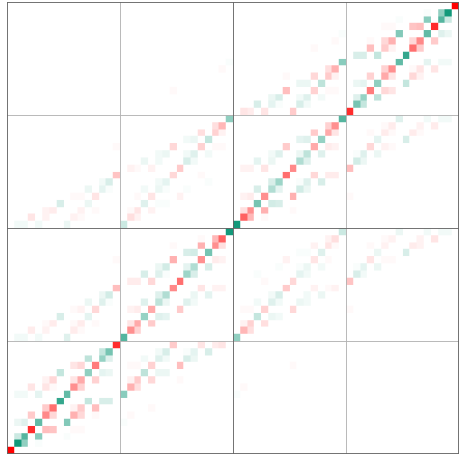,width=6cm}
\end{center}
\caption{\label{fig.heis.obc} Ground state of the 1D spin-$1/2$ AF
  Heisenberg model with {\em open} boundary conditions and $N=12$
  spins. Notice that, as opposed to the case of figure \ref{fig.heis},
  the secondary diagonals have almost vanished.}
\end{figure}

\begin{figure}
\begin{center}
\rput(3.75,2.75){\psline{->}(0,0)(1,0)\psline{<-}(0,-2)(1,-2)
\rput(0.5,0.2){$\cal R$}\rput(0.5,-2.25){$\cal R$}}
\hbox to 8.5cm{\epsfig{file=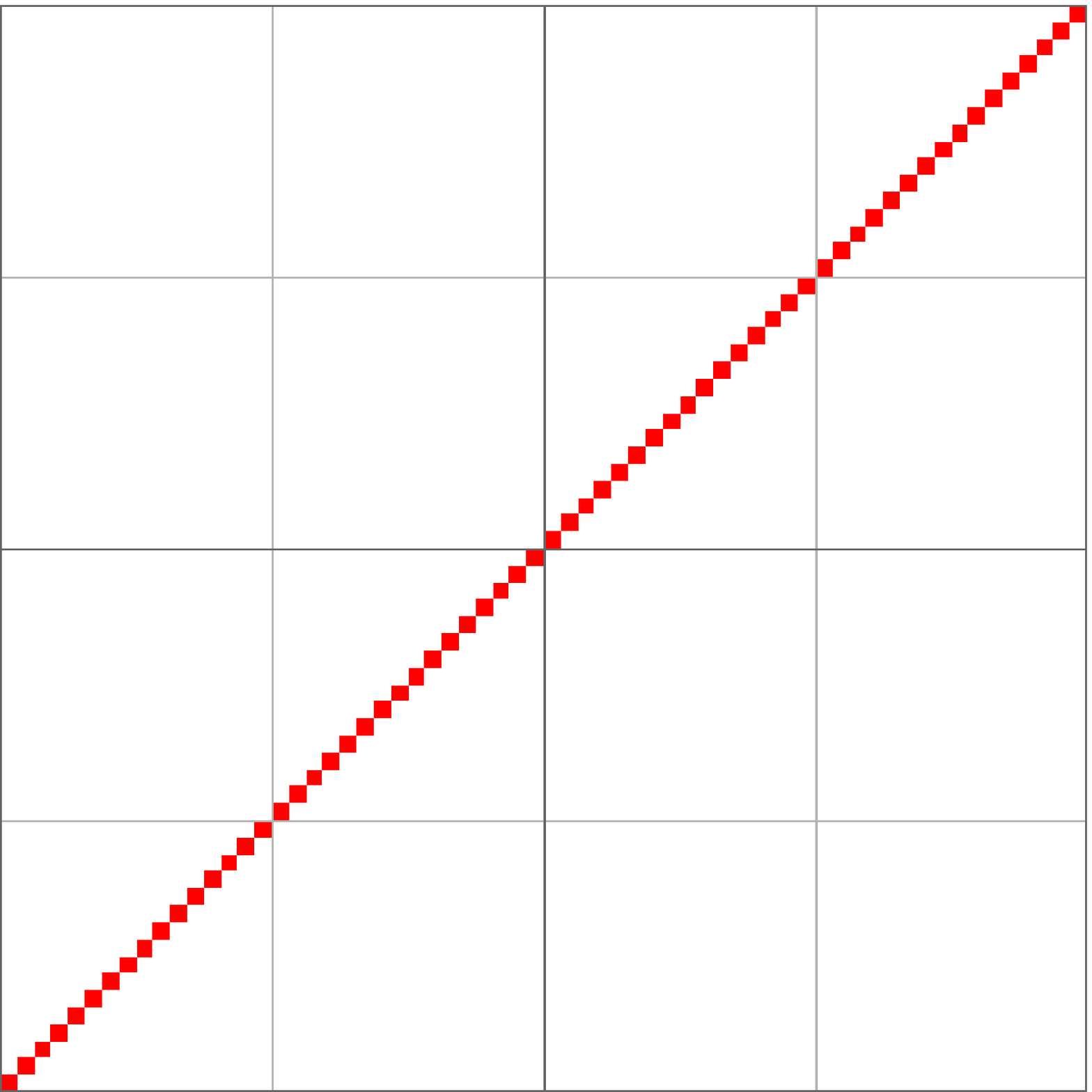,width=3.5cm}\hfill
\epsfig{file=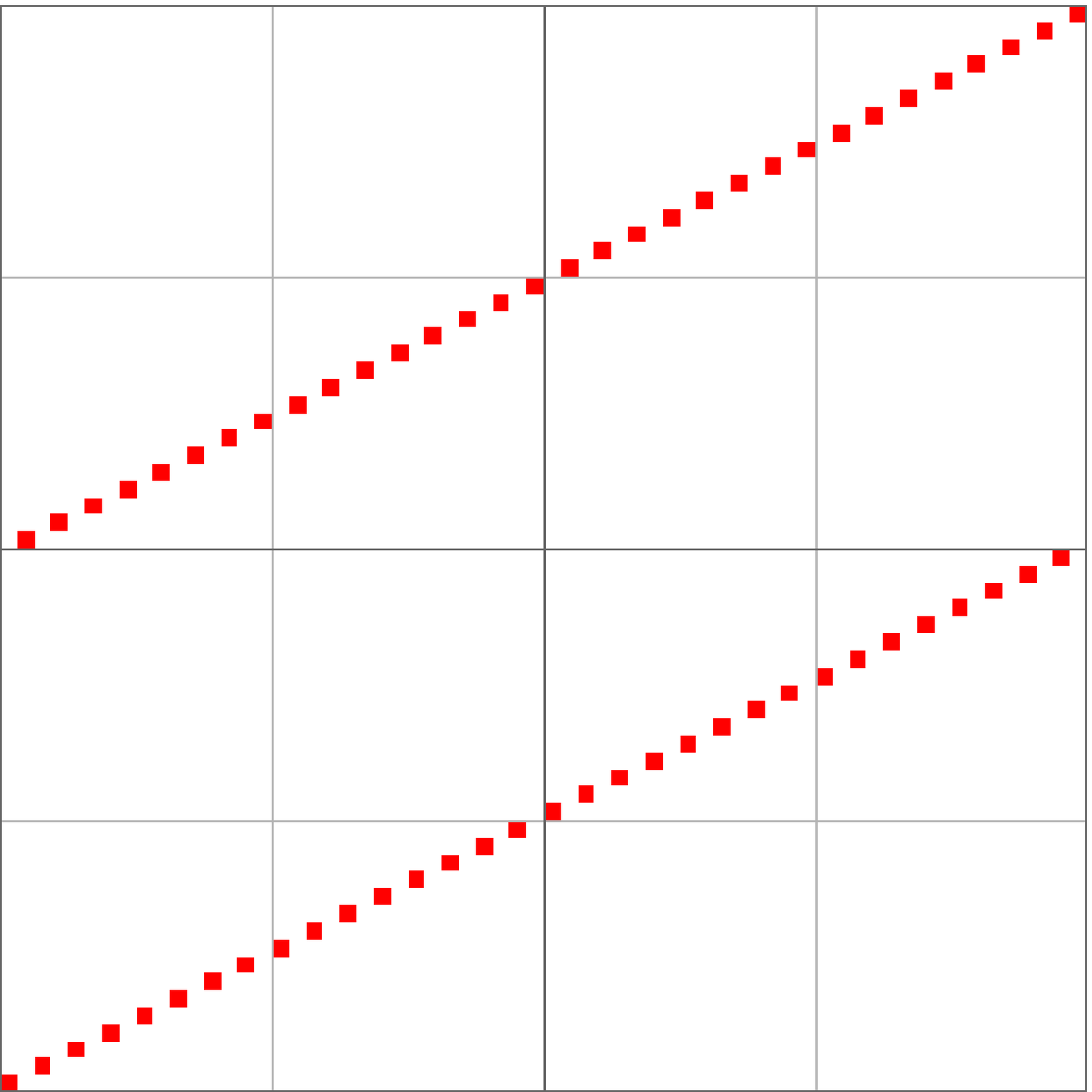,width=3.5cm}}
\end{center}
\caption{\label{fig.diagonals} The two diagonal structures which make
  up the Heisenberg ground state are related through a right-shift
  translation operator ${\cal R}$. If $(i,j)$ denotes a singlet bond,
  on the left we have $(1,2)(3,4)\cdots(N-1,N)$, and
  $(2,3)(4,5)\cdots(N,1)$ on the right.}
\end{figure}

There is another interesting feature in this image. The two parallel
diagonal lines with slope $1/2$ have the same intensity as the main
diagonal. What is their origin? A clue can be obtained when we depict
the GS of the Heisenberg model with {\em open} boundary conditions
(see figure \ref{fig.heis.obc}). It is apparent that these secondary
lines have almost disappeared. In order to finally clarify the nature
of these secondary lines, let us consider ${\cal R}$, the right-shift
translation operator (with periodic boundary conditions). If ${\cal
  R}$ acts on the states composing the main diagonal, the result is
the two secondary diagonals, and viceversa, as it can be seen in
figure \ref{fig.diagonals}. It is now straightforward to provide a
physical interpretation: the secondary diagonals depict the other
possible set of {\em pairwise singlet bonds}:
$(2,3)(4,5)\cdots(N,1)$. When periodic boundary conditions are
employed, both structures are equally important, but not under open
ones.

The slope $1/2$ of those secondary diagonals can be understood as
follows. According to equation \ref{coord.2d}, acting with the
right-shift translation operator ${\cal R}$ on a state given by bits
$\{X_1Y_1X_2Y_2\cdots X_nY_n\}$ we obtain $\{Y_nX_1Y_1X_2\cdots
Y_{n-1}X_n\}$. Thus, ${\cal R}$ maps the point $(x,y)$ into a point
very close to $((y+Y_n)/2,x)$. Consequently, the image of the $x=y$
line is approximately $x=(y+Y_n)/2$, i.e.: the two secondary lines. A
second application of the right-shift operator ${\cal R}$ on these two
secondary lines returns the original main diagonal. Of course, the
same effect is obtained with a left-shift.

\subsection{Next-nearest-neighbour Heisenberg: Marshall rule and Frustration}

Still there is one more interesting feature of the image of the ground
state of the Heisenberg Hamiltonian, \ref{fig.heis}. According to
Marshall's rule \cite{marshall.55}, the sign of each wavefunction
component of the ground state of the Heisenberg AF model in a
bipartite lattice (split into sublattices $A$ and $B$) can be given as
$(-1)^{N_A}$, where $N_A$ is the number of up-spins in sublattice
$A$. In our case, a 1D lattice with PBC, the two sublattices are just
the odd and even sites. It is not hard to recognize that, if we select
the odd sites to make up sublattice $A$, then the sign rule tells us
that all states in the same {\em horizontal} line must have the same
sign. But, on the other hand, if sublattice $A$ is made up with the
even sites, then the rule tells us that all states in the same {\em
  vertical} line will have the same sign. Both conditions can be
fulfilled, both in the PBC and the OBC figures, \ref{fig.heis} and
\ref{fig.heis.obc}.

\begin{figure}
\begin{center}
\begin{tabular}{cc}
\epsfig{file=H.012.001.pbc.eps,width=4cm} & 
\epsfig{file=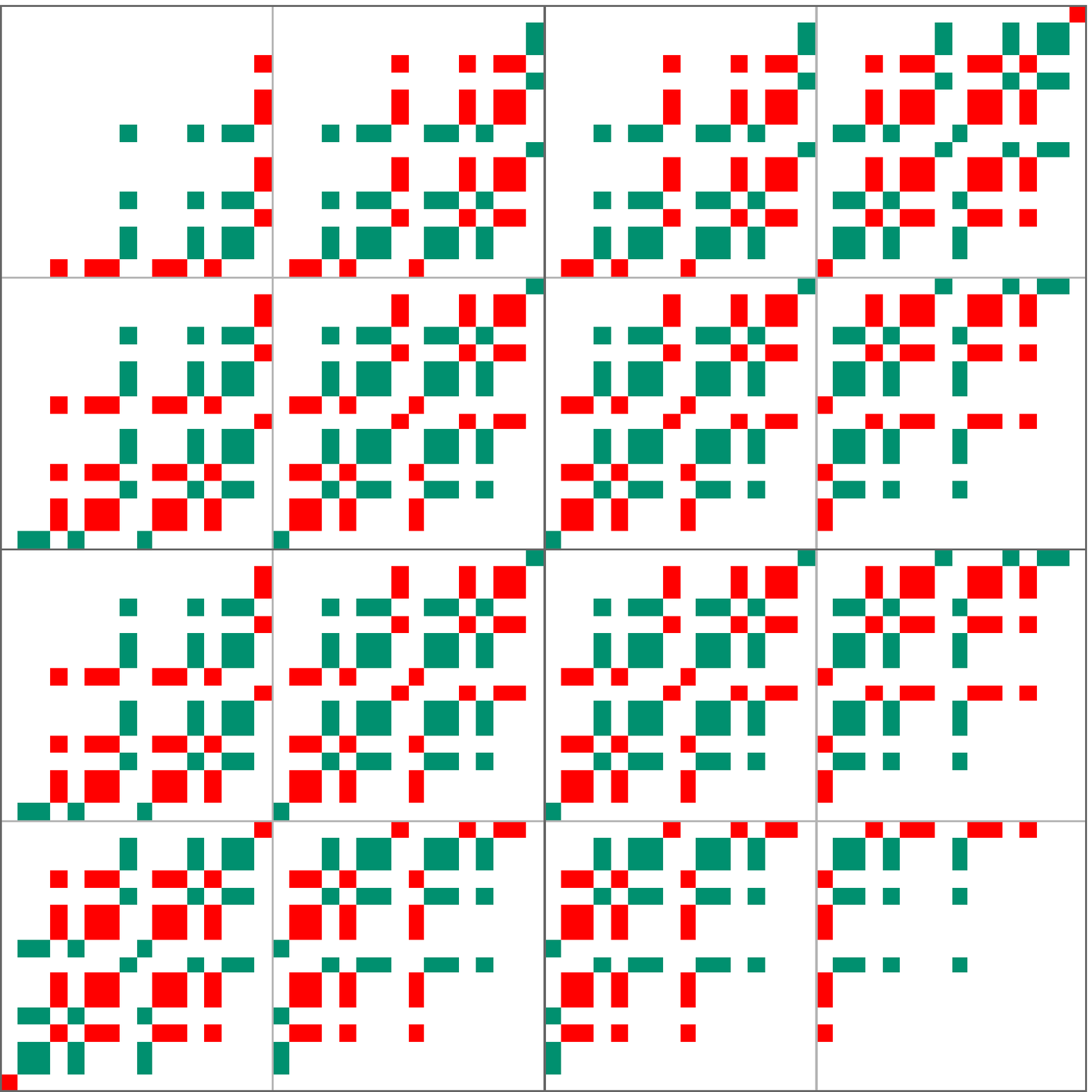,width=4cm} \\
\epsfig{file=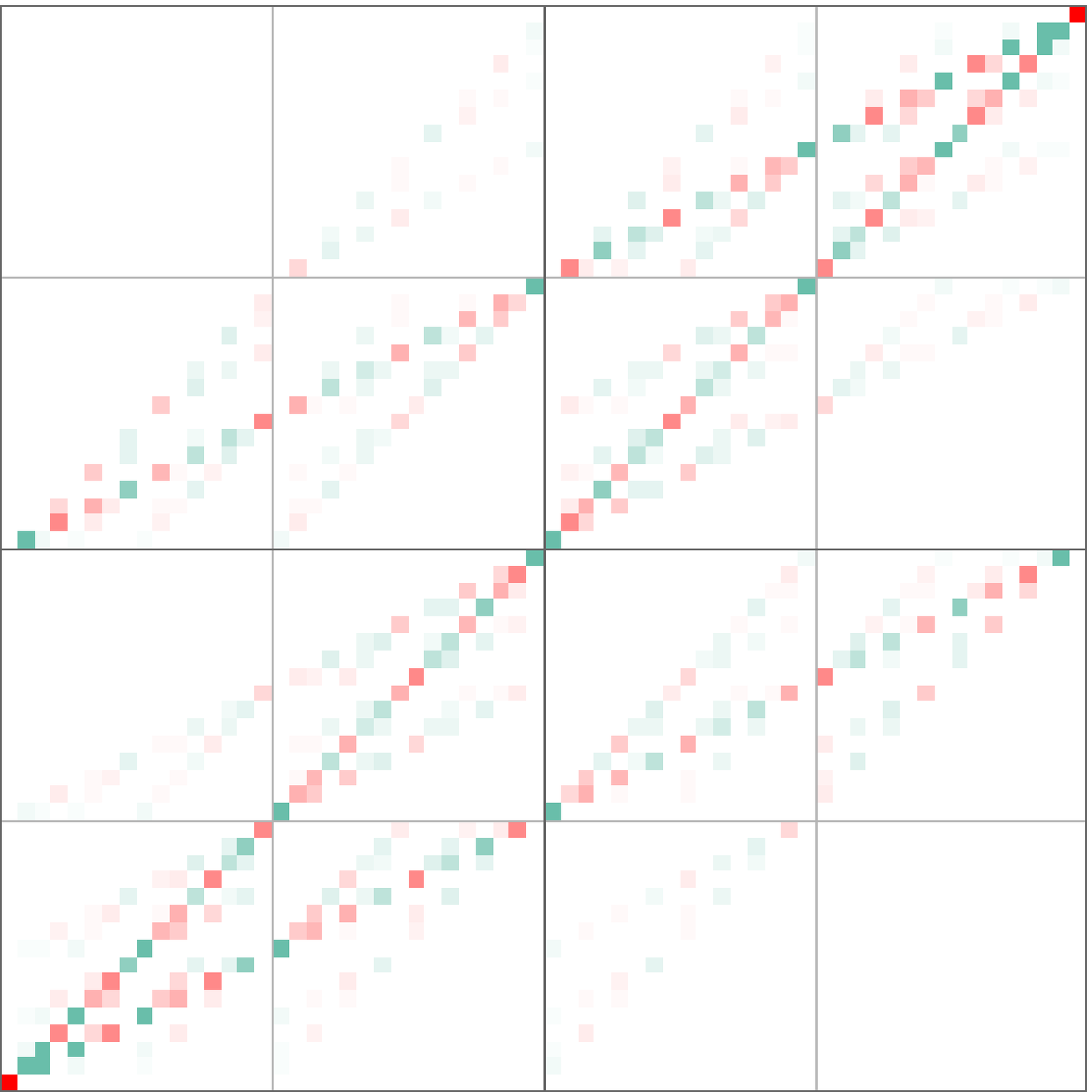,width=4cm} & 
\epsfig{file=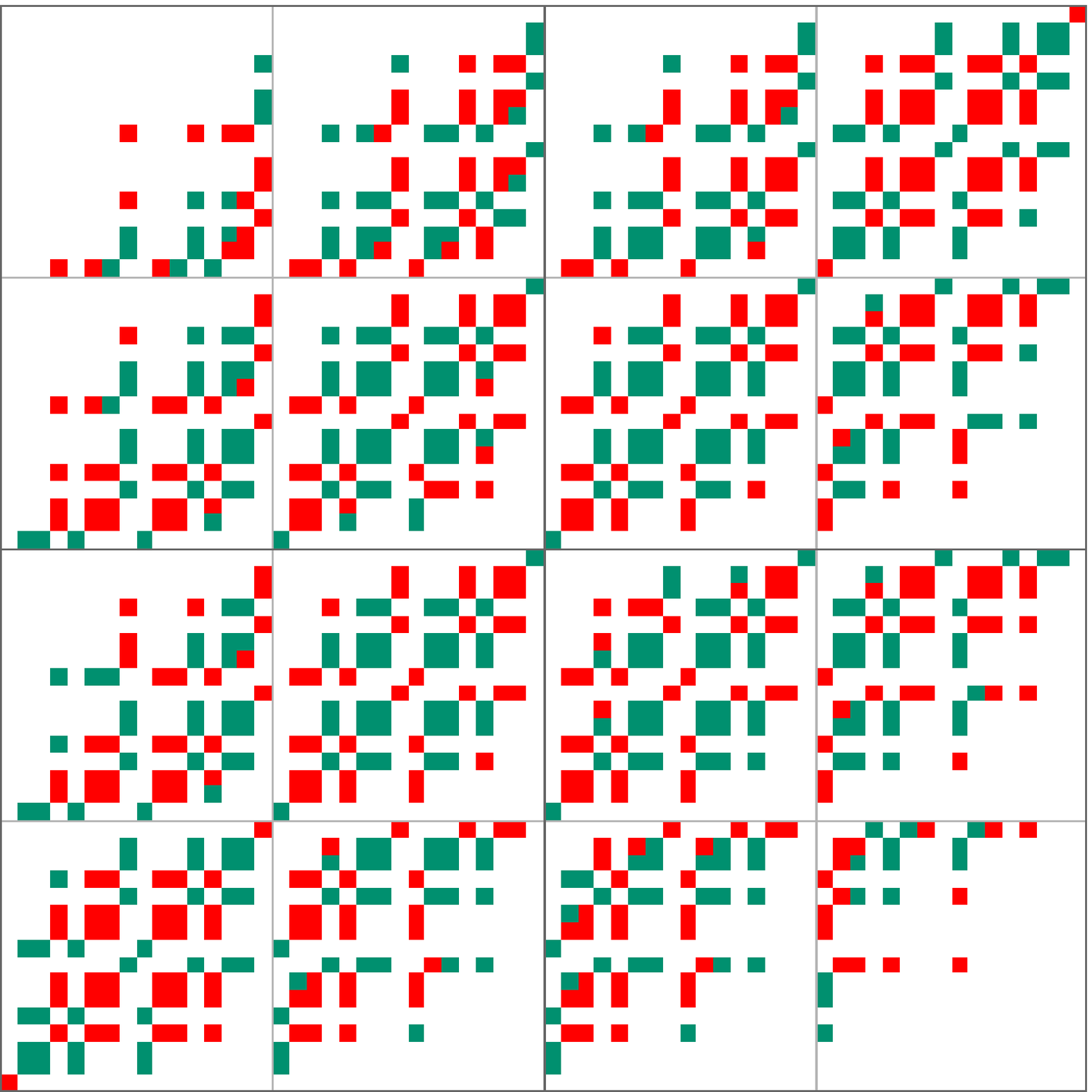,width=4cm} \\
\epsfig{file=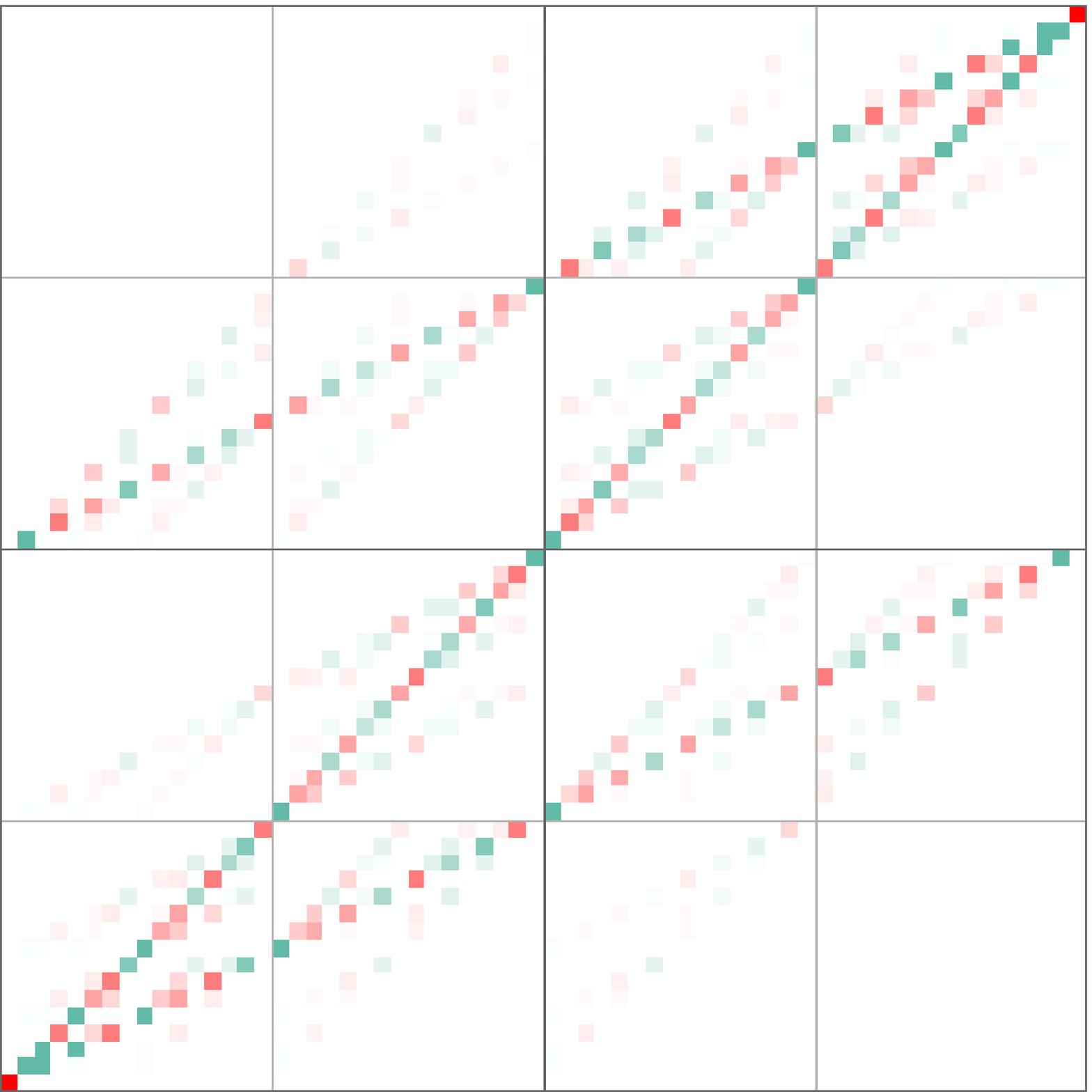,width=4cm} & 
\epsfig{file=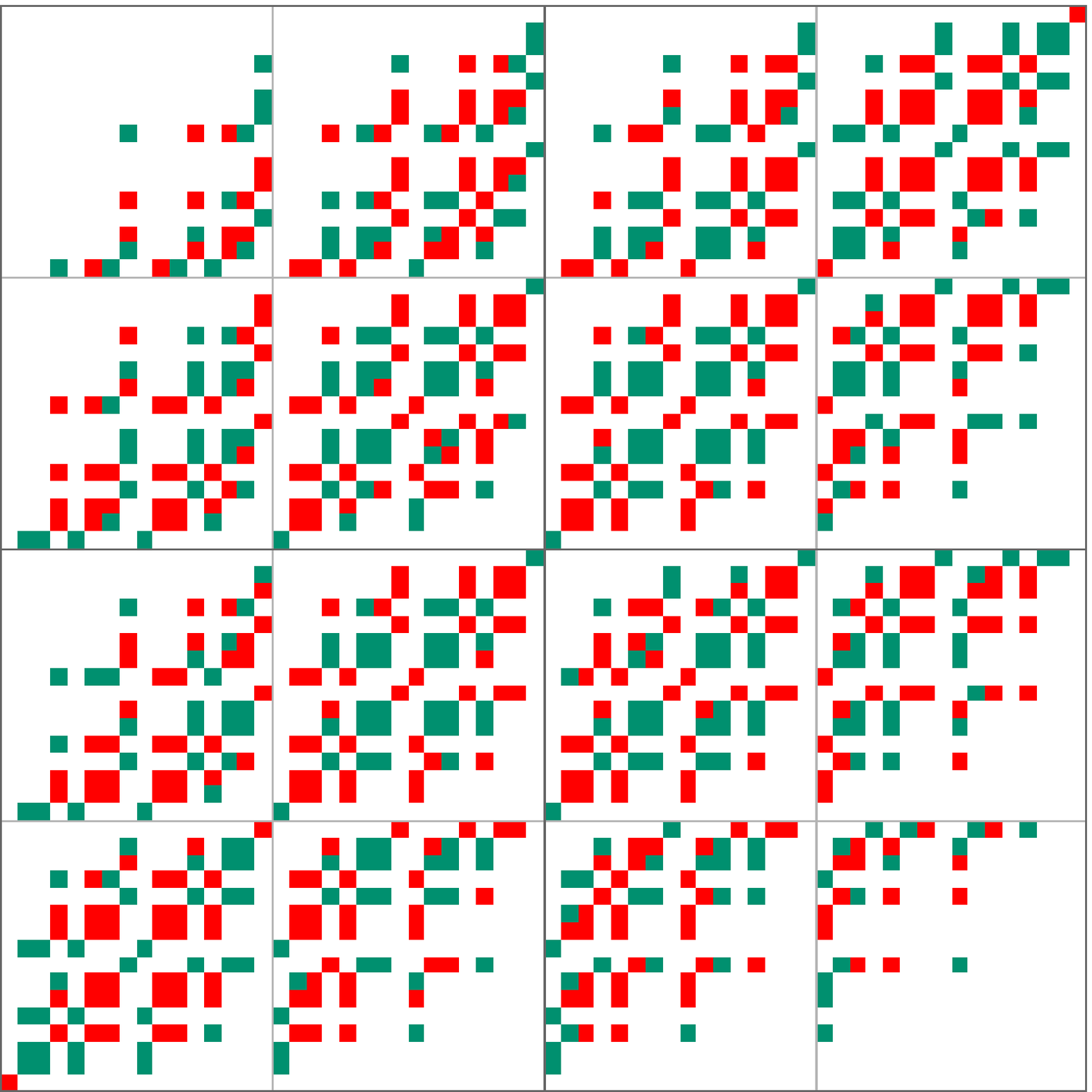,width=4cm} \\
\end{tabular}
\end{center}
\caption{\label{fig.j1j2} Destruction of the sign pattern given by
  Marshall rule as $J_2$ increases for the ground state of the
  Hamiltonian given in eq. \ref{j1j2.model}., with $N=12$ qubits and
  PBC. From top to bottom, the values of $J_2$ are $J_2=0$, $0.2$
  and $0.3$. The left column shows the full image. The right one only
  depicts the phases of the non-zero amplitude values.}
\end{figure}

\begin{figure}
\begin{center}
\epsfig{file=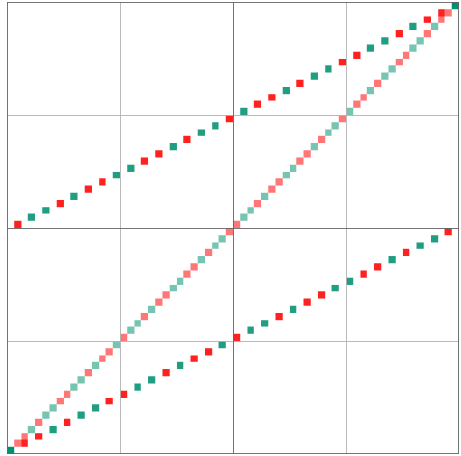,width=6cm}
\end{center}
\caption{\label{fig.mg}Majumdar-Ghosh state, ground state of the
  Hamiltonian in equation \ref{j1j2.model} with $N=12$ qubits and PBC
  for $J_2=0.5$.}
\end{figure}

Marshall sign rule can not be applied if the system presents {\em
  frustration}, i.e.: when the Hamiltonian couples spins in the same
sublattice $A$ or $B$. Let us consider the next-nearest-neighbour AF
Heisenberg Hamiltonian (also known as $J_1J_2$ model):

\begin{equation}
H=J_1\sum_{i=1}^N {\vec S}_i \cdot {\vec S}_{i+1} + J_2 \sum_{i=1}^N
{\vec S}_i \cdot {\vec S}_{i+2}
\label{j1j2.model}
\end{equation}

\noindent where $J_1=1$ and $J_2>0$. Then, as $J_2$ increases, the
system undergoes a quantum phase transition (QPT) at around
$J_2\approx 0.24$. Figure \ref{fig.j1j2} shows how the sign-structure
is destroyed slowly as $J_2$ is increased from $J_2=0$ to
$J_2=1/2$. The point $J_2=1/2$ is special, since the ground state is
then exactly known: the Majumdar-Ghosh point. Its rather simple
structure is apparent in figure \ref{fig.mg}.

\subsection{\label{section.factorizable}Product States}

Let us now consider the simplest possible quantum many-body
wavefunction: a translationally invariant {\em product state},
defined as

\begin{equation}
\ket|\Psi>=\( \alpha\ket|0>+\beta\ket|1> \)^N
\label{factorizable}
\end{equation}

\begin{figure}
\begin{center}
\epsfig{file=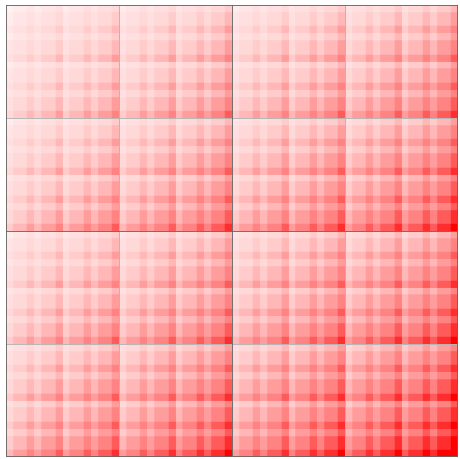,width=6cm}
\end{center}
\caption{\label{fig.fact}Product state following equation
  \ref{factorizable} with $\beta/\alpha=1.3$.}
\end{figure}

Factorizability is a very strong property, which shows itself in a
very appealing way in our plots. Figure \ref{fig.fact} shows such a
product state in the $\sigma_z$ basis. Physically, factorizability
implies that measurements performed on any qubit should have no
influence on the remaining ones. Concretely, we can measure $\sigma_z$
on the first two qubits. If the result is $00$, the wavefunction-plot
which describes the rest of the system will be (a normalized and
rescaled version of) the upper-left quadrant of the
plot. Correspondingly, if the results are $01$, $10$ or $11$, the
wavefunction-plot will be just a (normalized and rescaled version of)
other quadrant. Thus, factorizability implies that {\em all four
  quadrants are equal} (modulo a normalization factor). This line of
thought can be extended to the set of the first $2k$ qubits, thus
showing that if we split the plot into a $2^k\times 2^k$ array of
sub-images, they should all coincide, modulo a normalization
factor. This gives the characteristic look to the plots of product
states. We will return to this topic in section \ref{entanglement},
when we discuss entanglement.

\subsection{Dicke states}

\begin{figure}
\begin{center}
\begin{tabular}{cc}
\epsfig{file=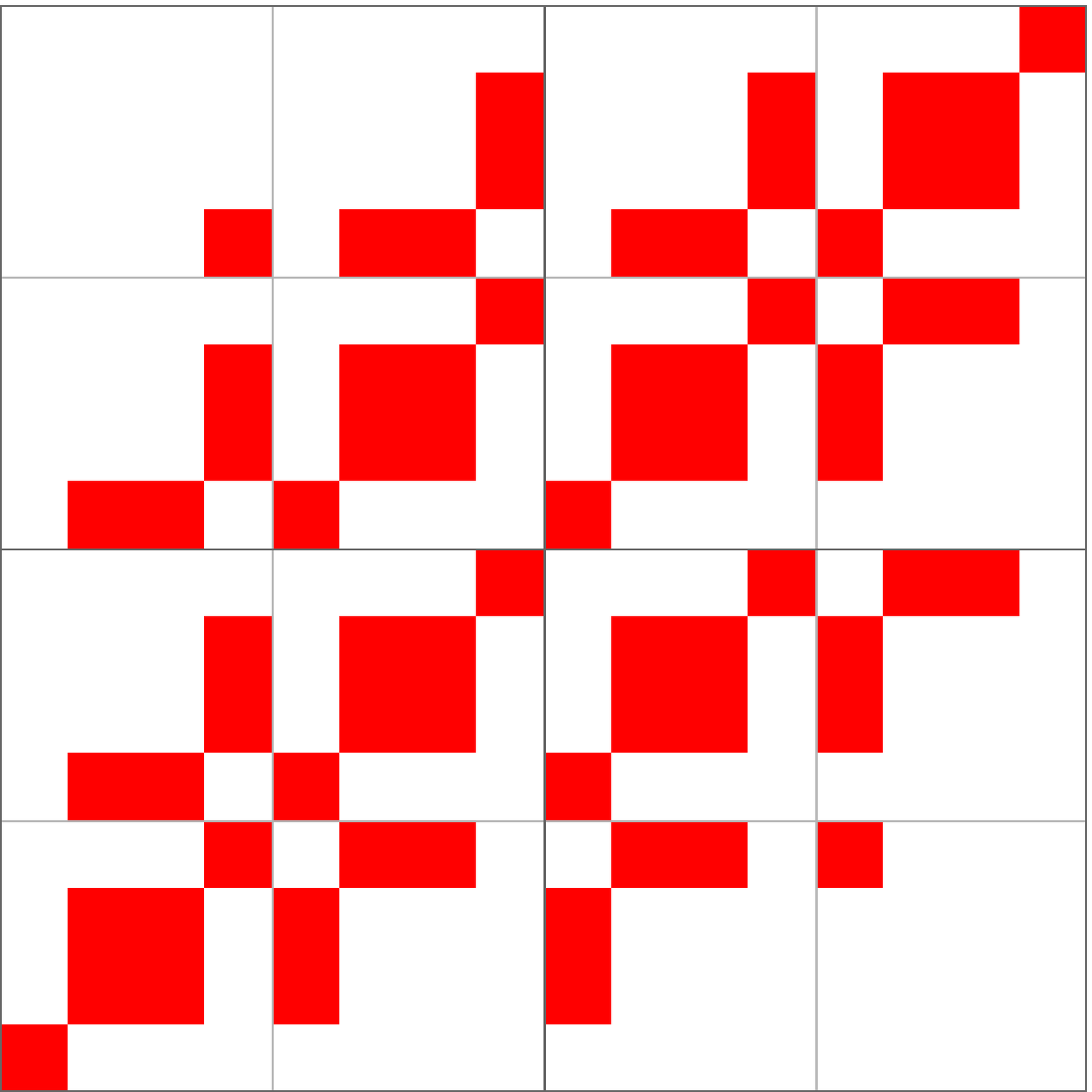,width=4cm} &
\epsfig{file=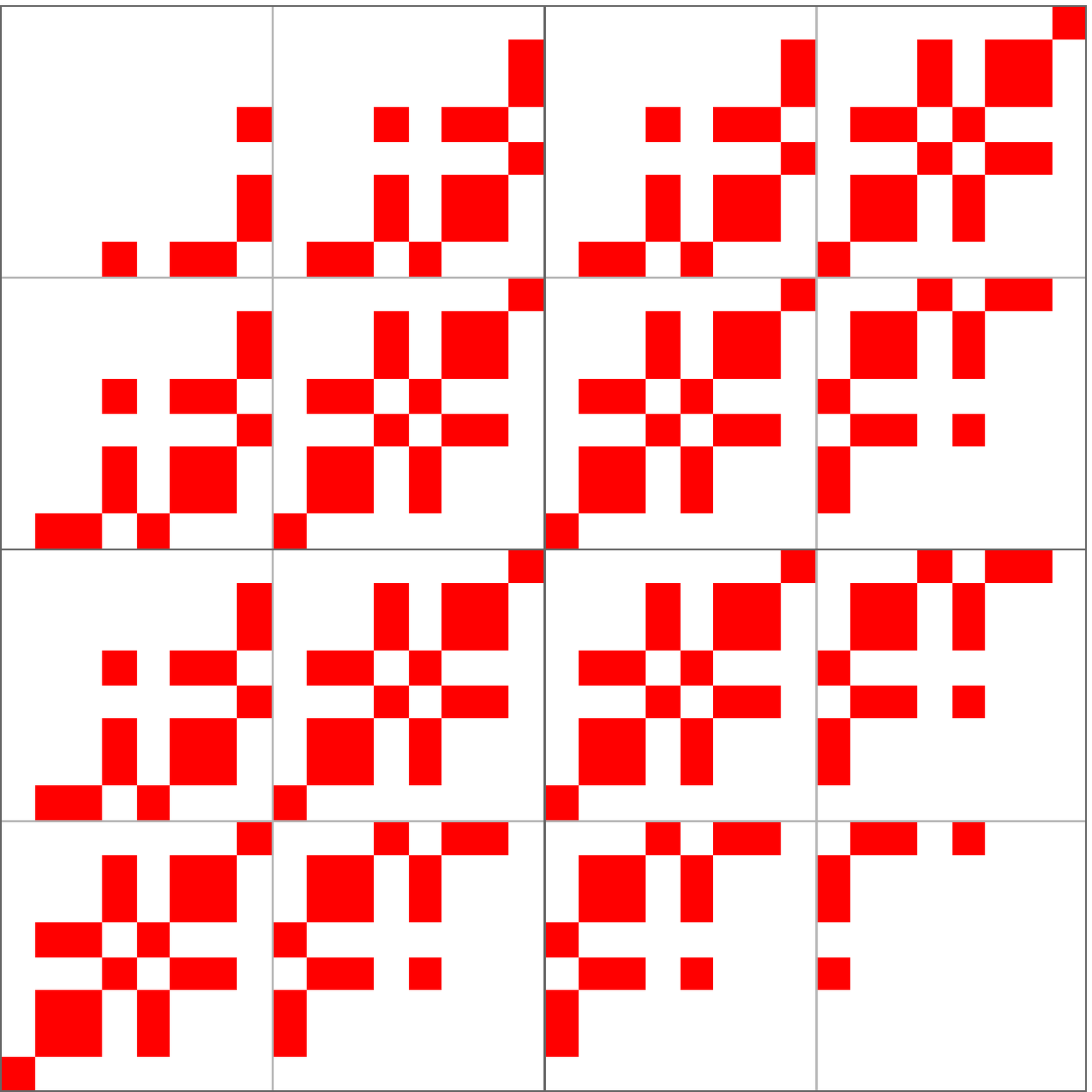,width=4cm} \\
\epsfig{file=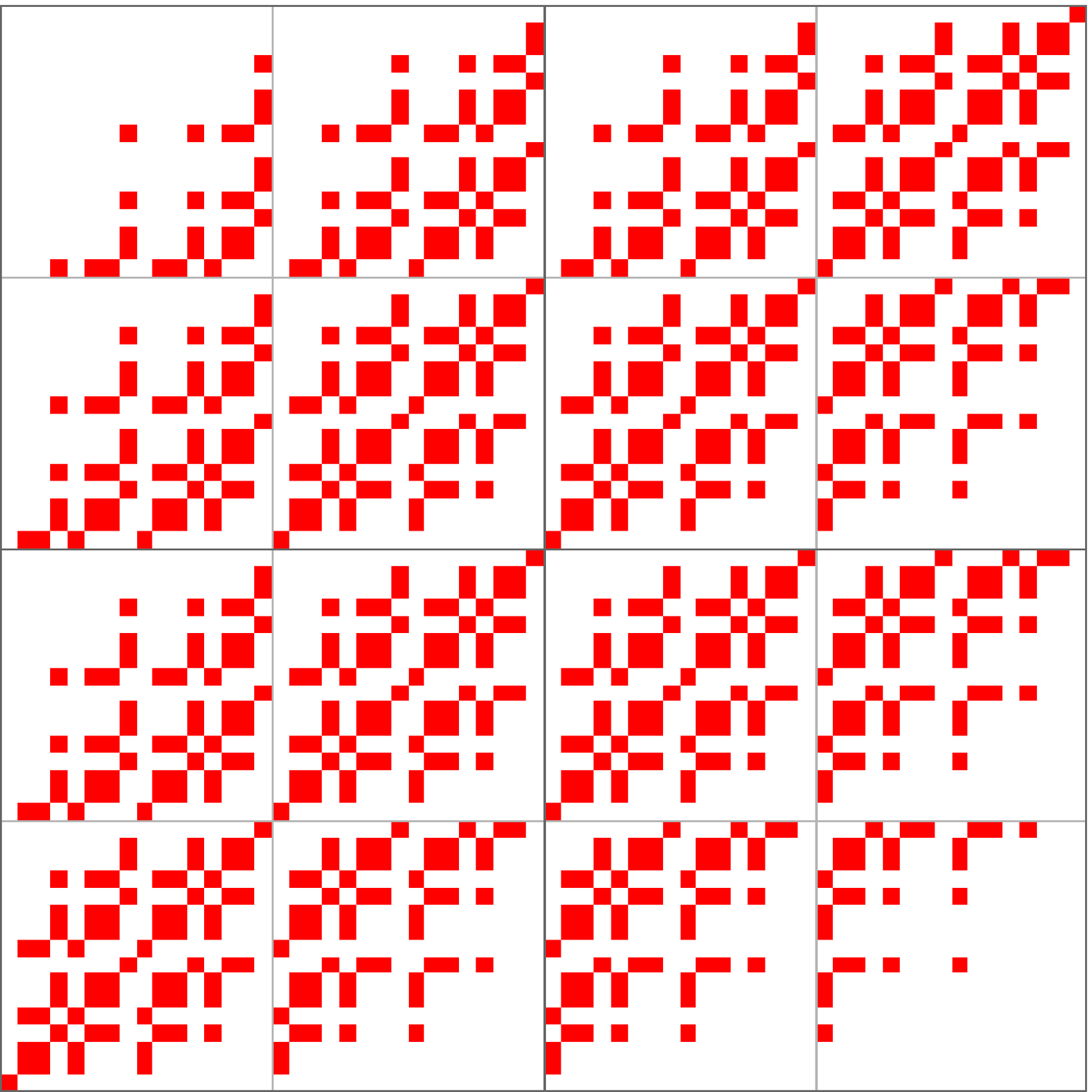,width=4cm} &
\epsfig{file=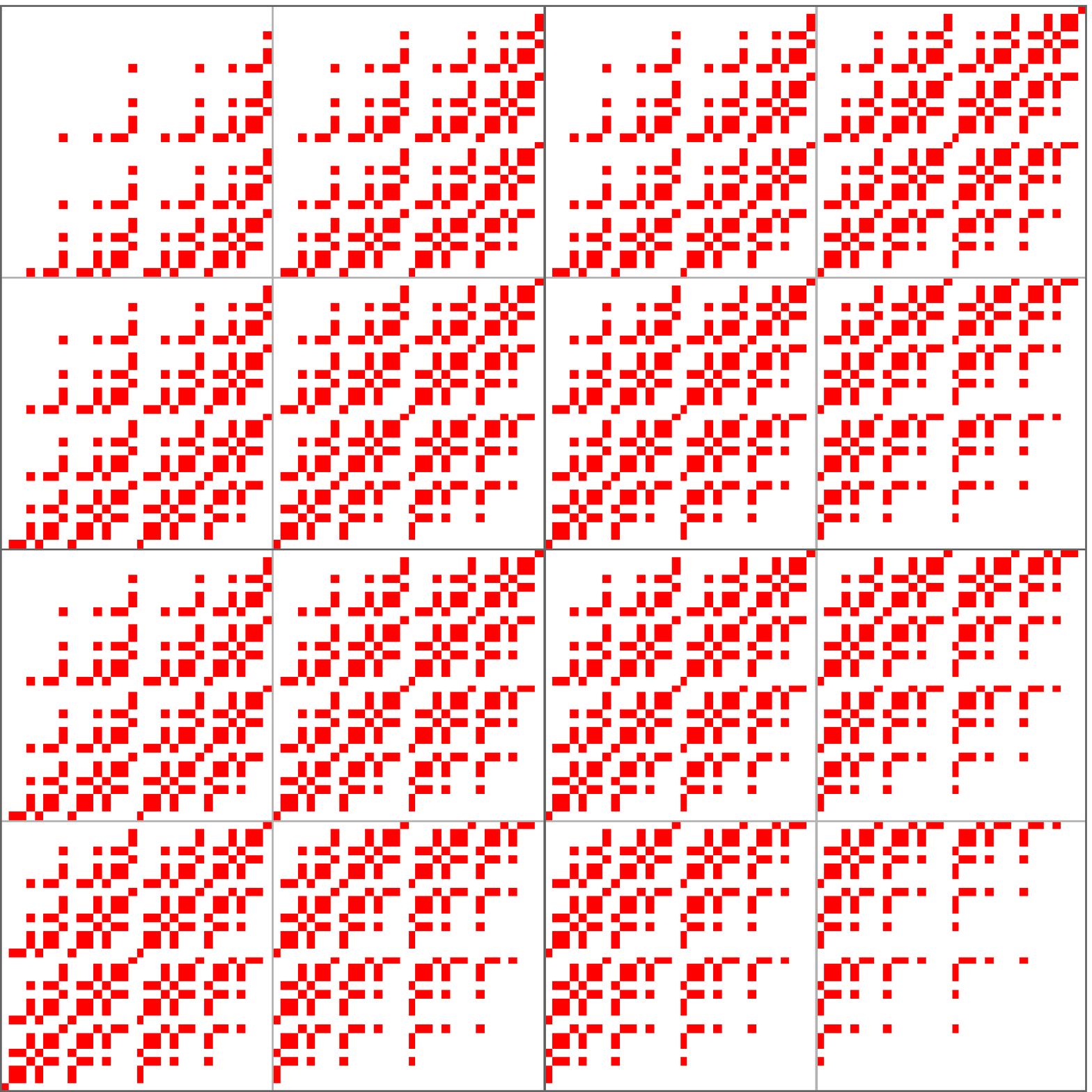,width=4cm} \\
\end{tabular}
\end{center}
\caption{\label{fig.dicke} Half-filling Dicke states for $N=8$, $10$,
  $12$ and $14$ qubits. Notice how the fractal structure develops.}
\end{figure}

Another interesting example is provided by the so-called Dicke states
\cite{stockton.03}. Those are defined as the linear combination, with
equal weights, of all tensor basis states with the same number $n_e$
of $1$'s in their decomposition. In our examples, we will focus on the
half-filling case, $n_e=N/2$. In fermionic language, they constitute
the ground state of a free-fermion model with homogeneous diffusion on
a complete graph at half-filling, and in spin-$1/2$ language it is the
$S_z=0$ component of the maximal spin multiplet. Figure
\ref{fig.dicke} shows the pattern obtained for different lattice
sizes. It is apparent how a fractal develops. Their similarity to the
right column of figure \ref{fig.j1j2} is, of course, not casual: the
ground states of the Heisenberg-like models have global magnetization
zero, which make them similar to half-filling Dicke states.

\subsection{Ising model in a transverse field: Criticality}

As a different relevant example, let us consider the spin-$1/2$ AF
Ising model in a transverse field (ITF), in a 1D chain with PBC:

\begin{equation}
H=\sum_{i=1}^N \sigma^z_i \sigma^z_{i+1} - \Gamma \sum_{i=1}^N \sigma^x_i
\label{itf.model}
\end{equation}

\begin{figure}
\begin{center}
\begin{tabular}{cc}
\epsfig{file=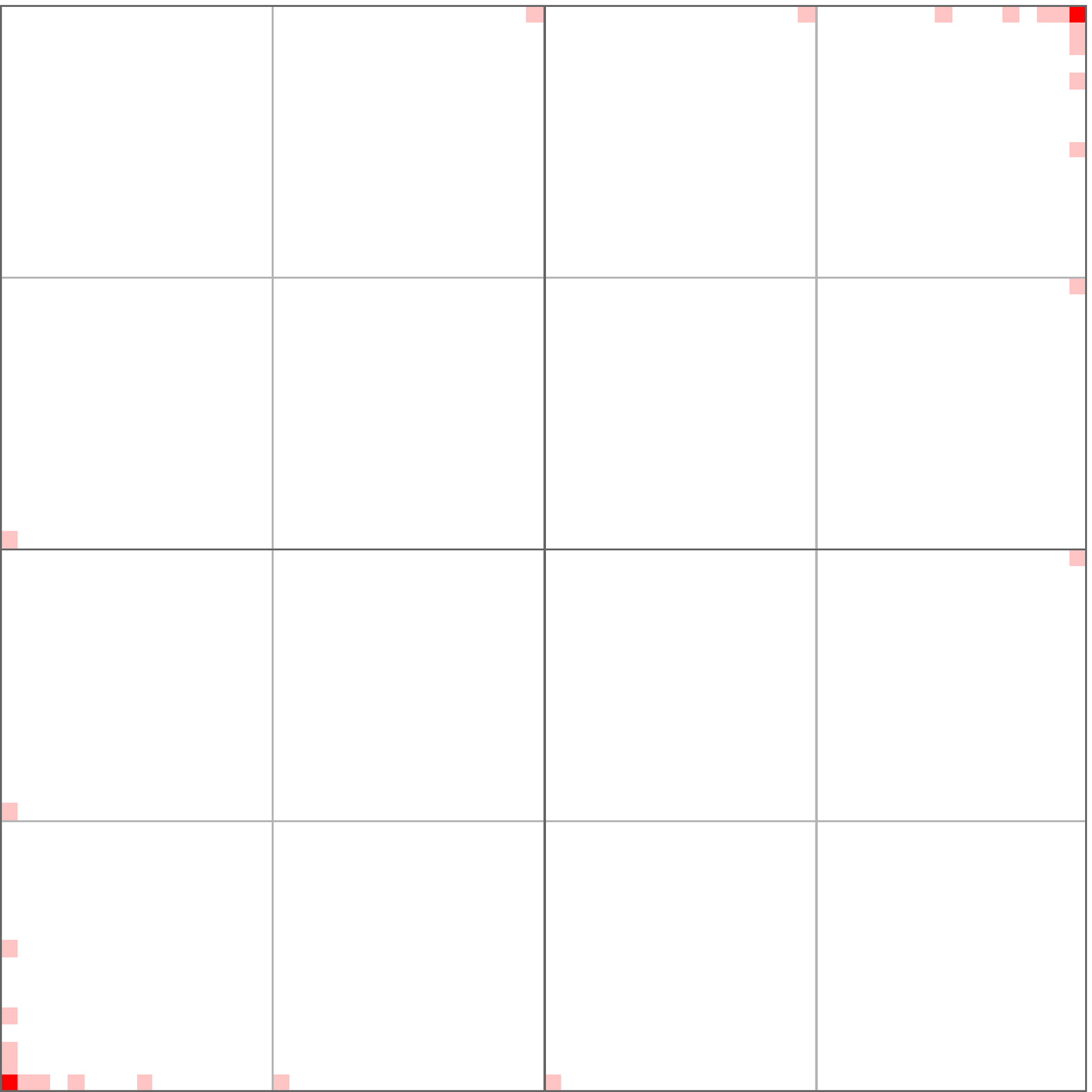,width=4cm} &
\epsfig{file=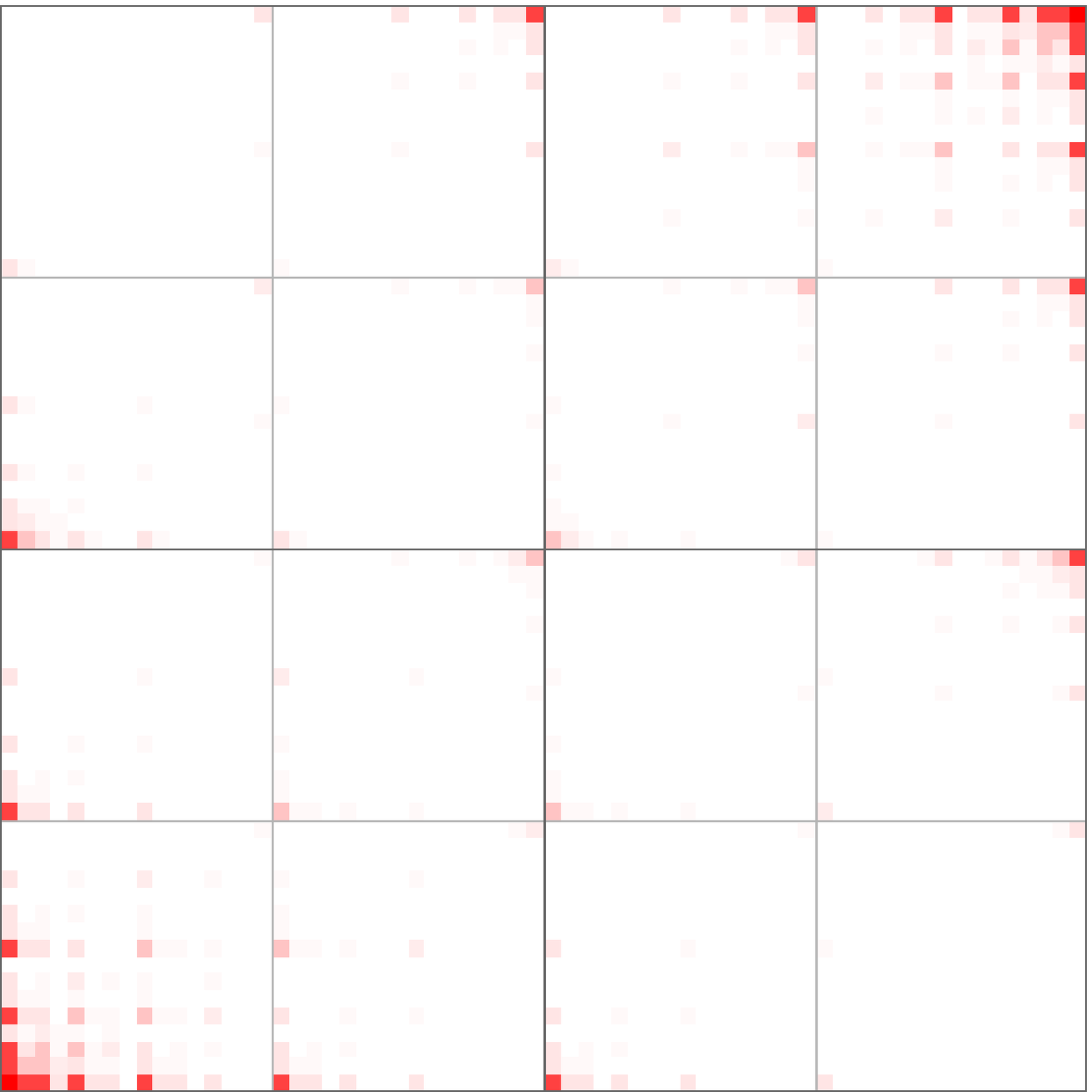,width=4cm} \\
\multicolumn{2}{c}{\epsfig{file=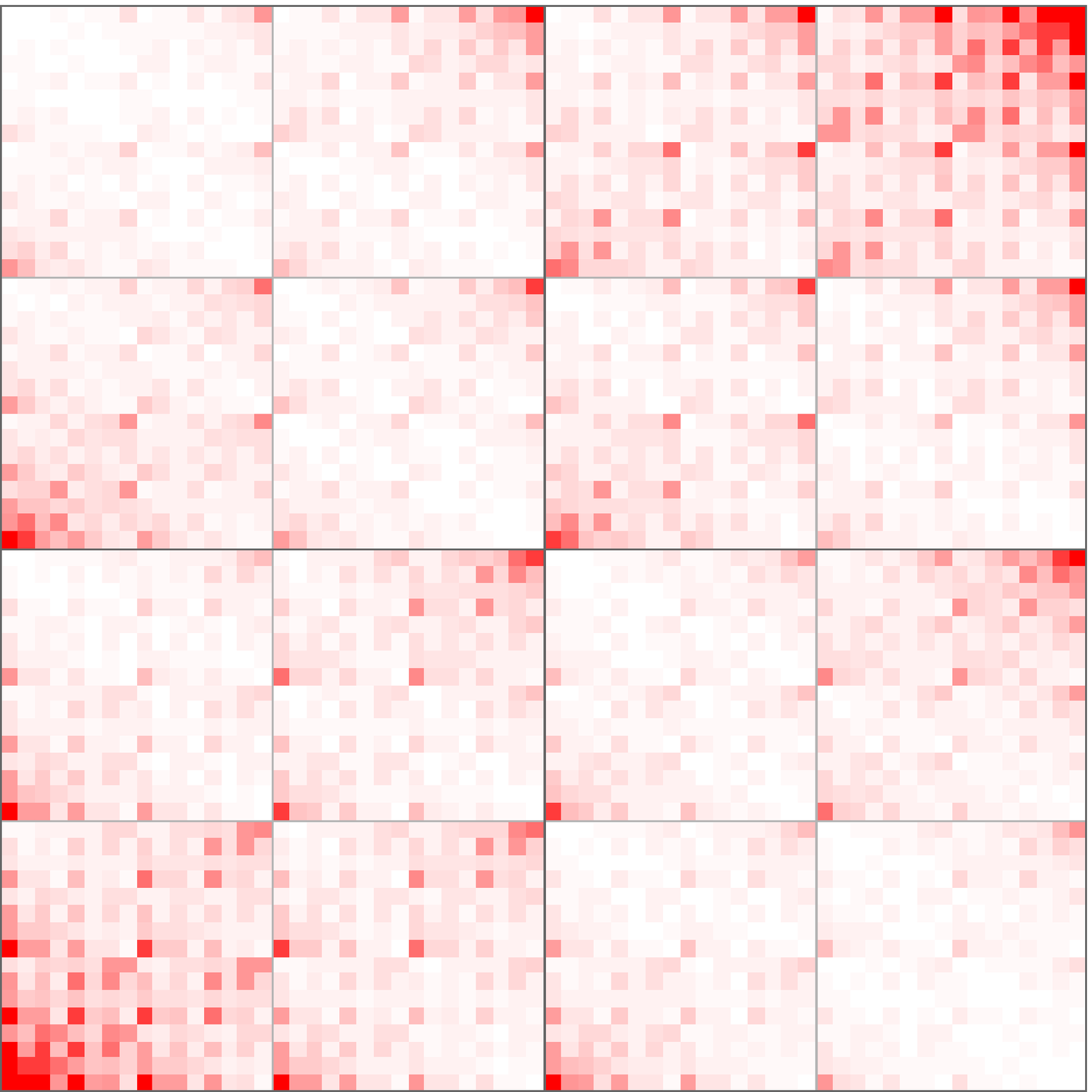,width=6cm}} \\
\epsfig{file=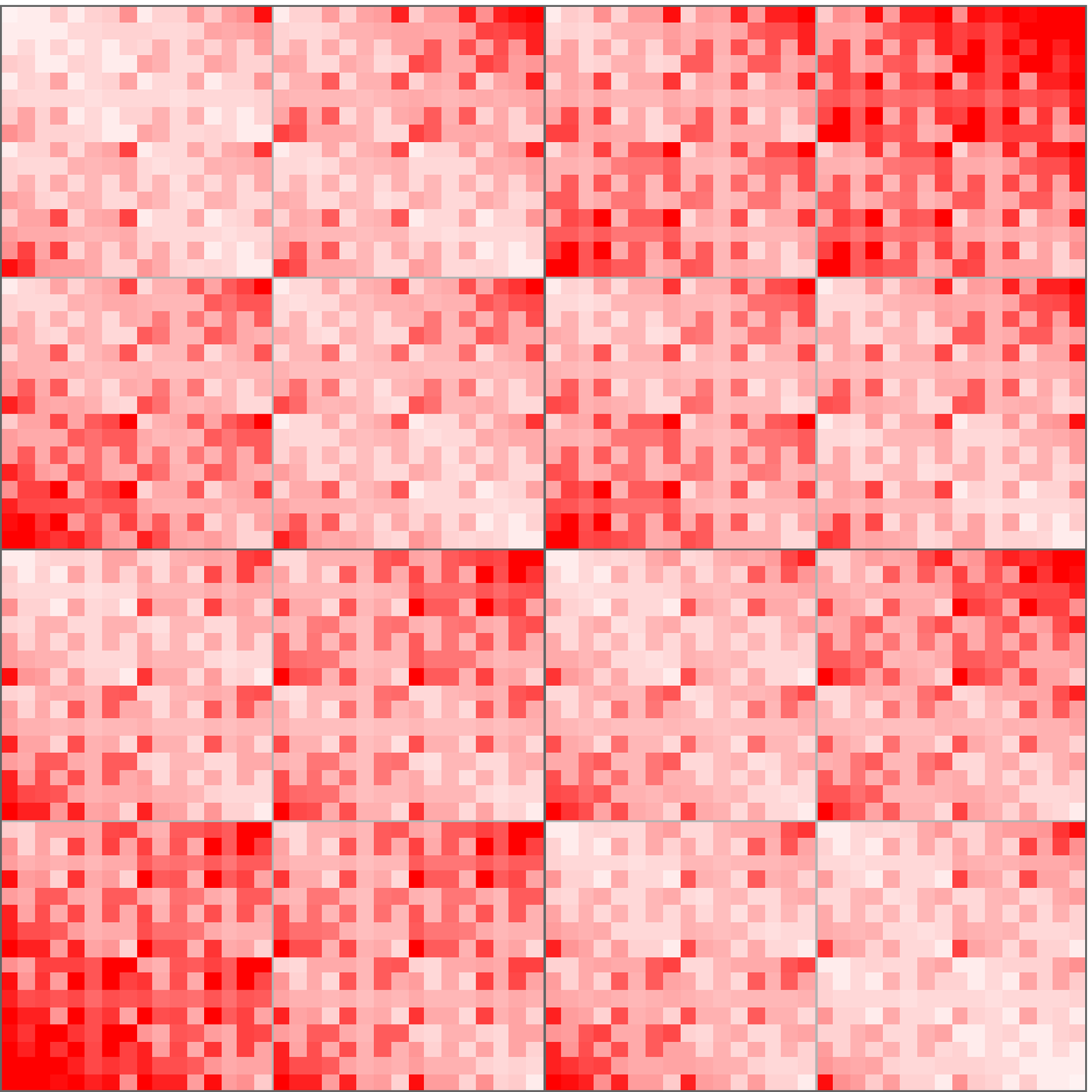,width=4cm} &
\epsfig{file=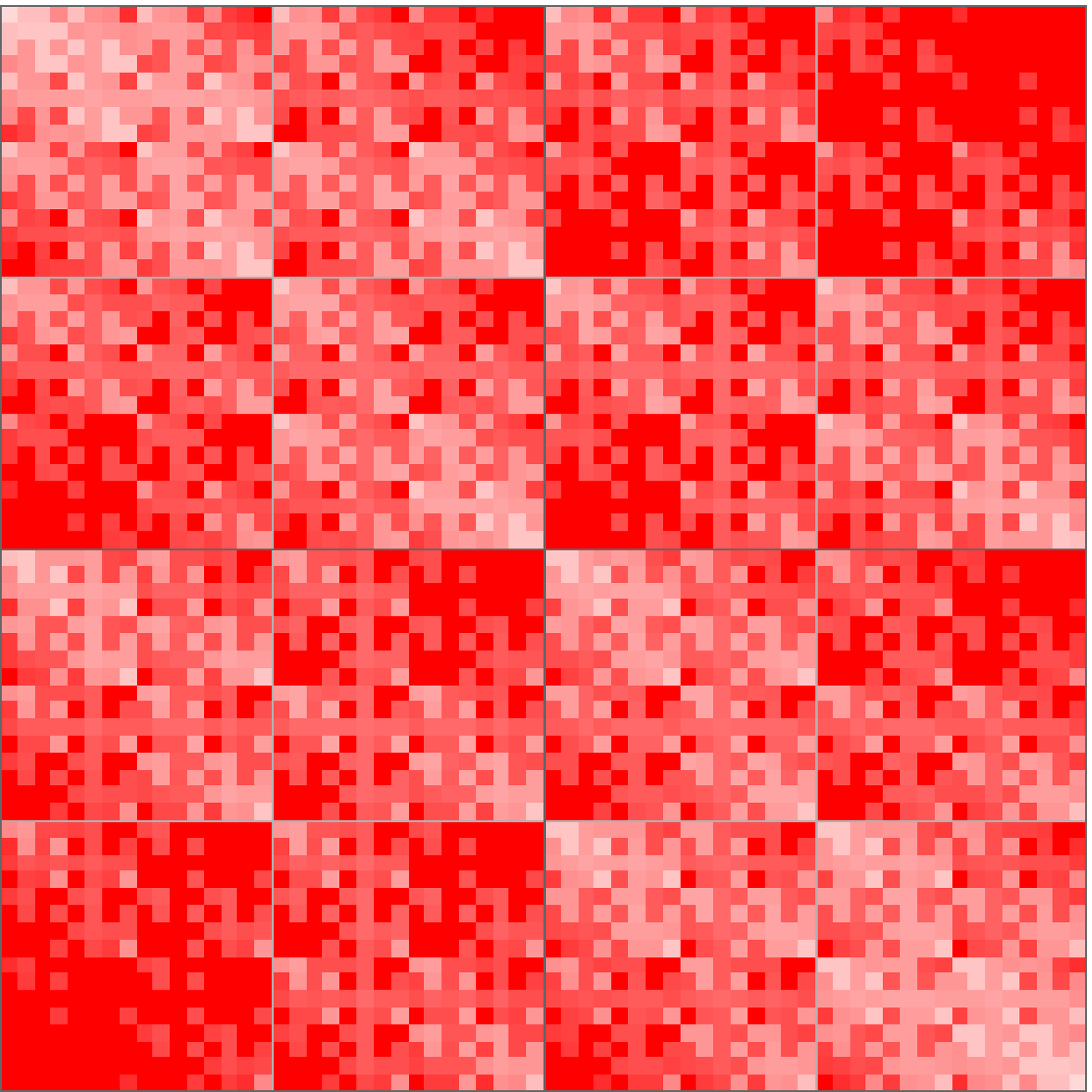,width=4cm} 
\end{tabular}
\end{center}
\caption{\label{fig.itf} Ground states of the AF ITF Hamiltonian with
  $N=12$ qubits and PBC. Values of the transverse field are: (top)
  $\Gamma=0.2$ and $0.6$, (central) $\Gamma=\Gamma_c=1$, (bottom)
  $\Gamma=1.4$ and $1.8$.}
\end{figure}

For $\Gamma=1$, the system presents a quantum phase transition (QPT).
Figure \ref{fig.itf} shows the plots obtained from the GS for
different values of $\Gamma$. For $\Gamma\to 0$, the ground state
consists only of the two N\'eel states. As $\Gamma$ increases (first
two top panels), the points which come up first correspond to a single
defect, at all possible positions in the lattice. The non-zero
probability amplitudes extend further away from the original corner
states as $\Gamma$ approaches criticality, and at that point
$\Gamma_c=1$, the non-zero values have extended through the whole
image, albeit quite inhomogeneously. From that point, increasing
$\Gamma$ makes the image more and more homogeneous. For
$\Gamma\to\infty$, the ground state would consist of all spins
pointing in the $X$-direction, and this implies that the wavefunction
components will all take the same value.

\subsection{Infinite-range Hamiltonians}

And let us finish this section by considering infinite-range
Hamiltonians, i.e.: those in which all spins are linked to all
others. They can be thought of as infinite-dimension or {\em
  mean-field} systems. Those Hamiltonians commute with the full set of
generators of the permutation group. Therefore, their ground states
are often invariant under it. Compared to translation invariance, this
symmetry group is so large ($N!$ elements vs. $N$) that it leaves very
little freedom: a fully permutation-invariant wavefunction of $N$
qubits is characterized by just $N+1$ independent components, one per
global magnetization sector. Thus, permutation-invariant wavefunctions
have a clear visual fingerprint.

\begin{figure}
\epsfig{file=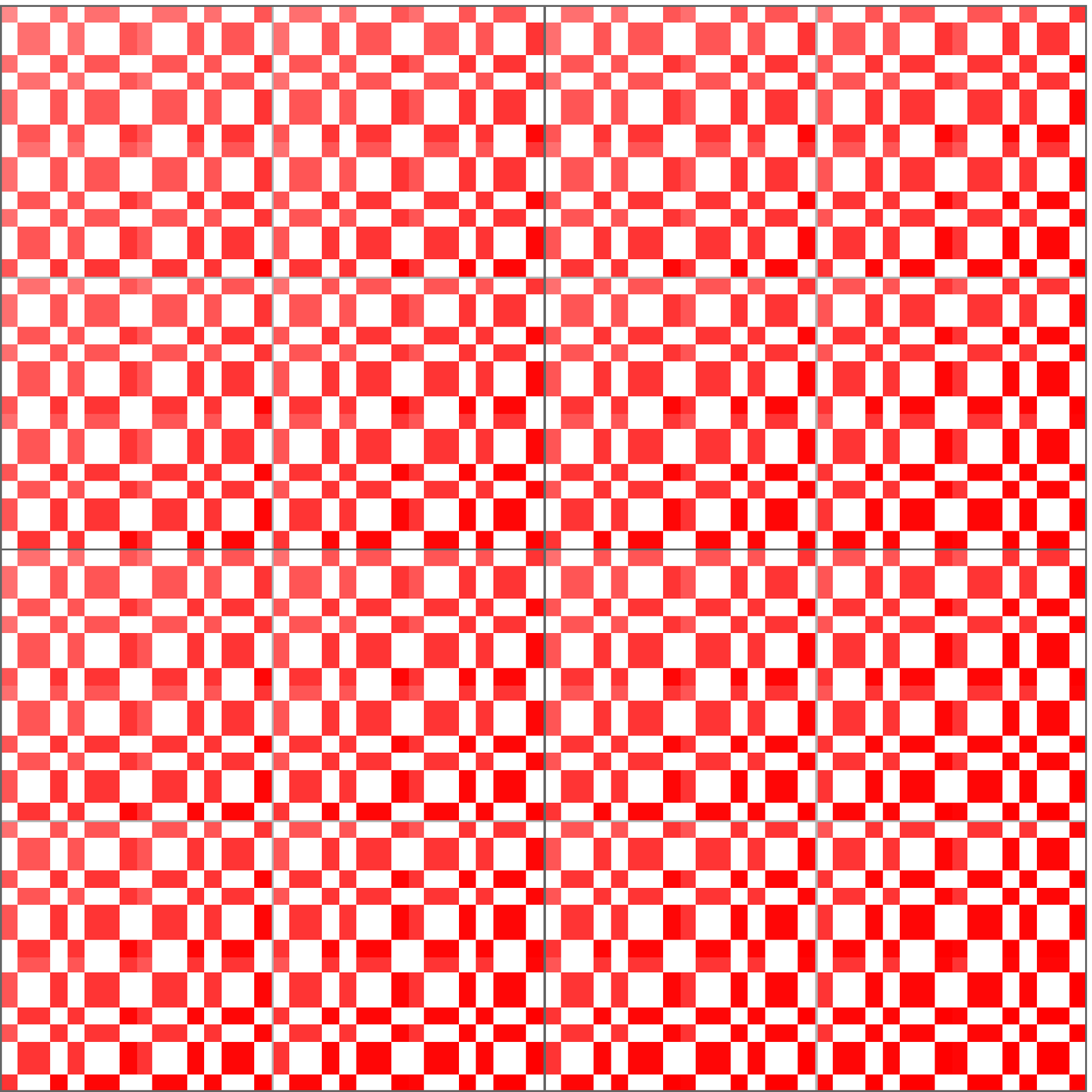,width=6cm}
\epsfig{file=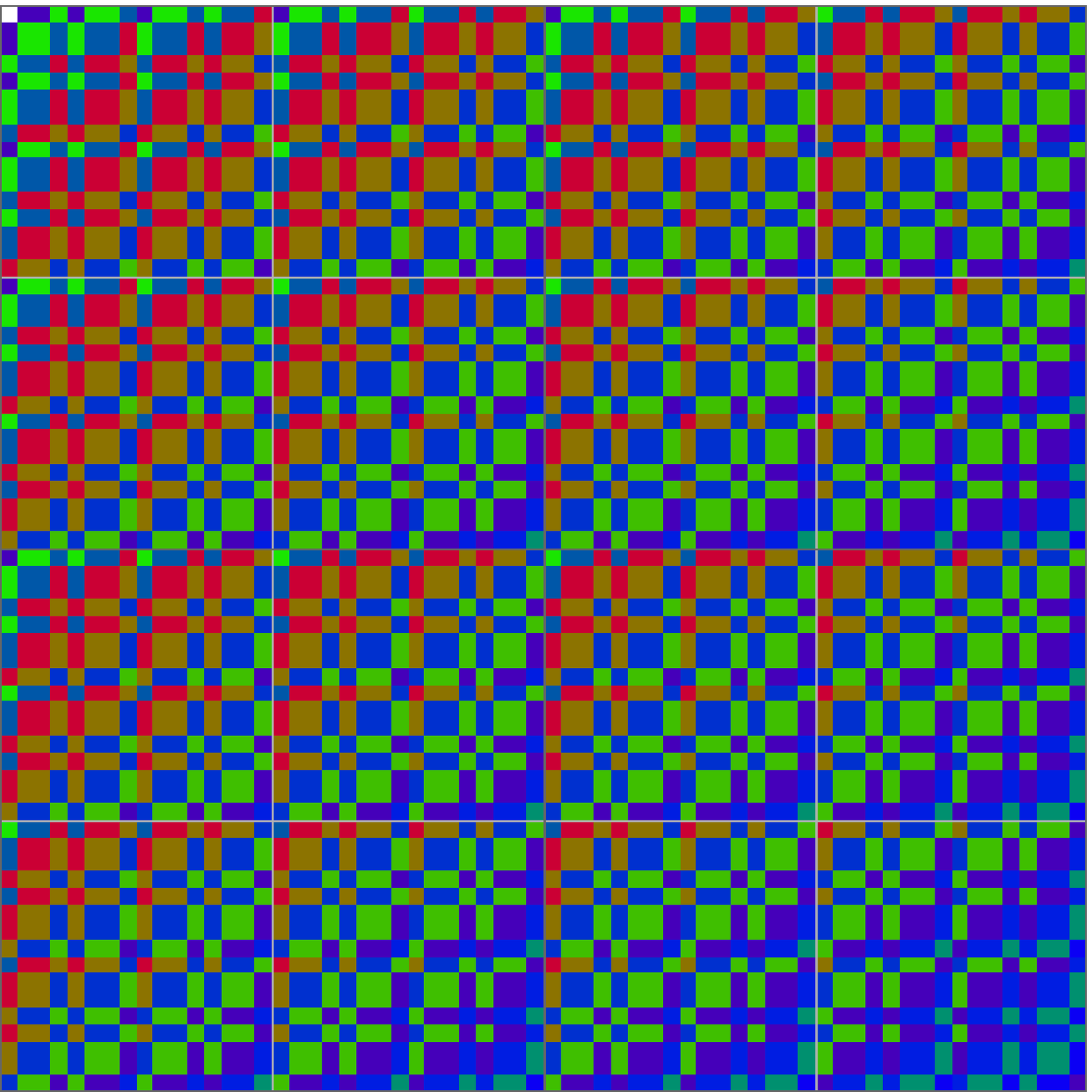,width=6cm}
\caption{\label{fig.mf} Left: Ground state of the AF infinite-range ITF
  model for $\Gamma=1$ with $N=12$ qubits. Right: a random
  permutation-invariant state, also with $N=12$ qubits.}
\end{figure}

Figure \ref{fig.mf} (left) shows the GS of the infinite-range AF ITF
Hamiltonian for $\Gamma=1$ and $N=12$ qubits, illustrating this high
degree of symmetry. The right part of the figure shows a random
permutation-invariant state. It is not a coincidence that it reminds
so strongly of the Dicke states, since each magnetization sector
shares the same color. The infinite-range Heisenberg Hamiltonian
ground state is not shown because it is strongly degenerate, so
invariance under the permutation group remains only as a property of
the full subspace.

\section{\label{otherplots}Other Plotting Schemes}

\subsection{General Formulation}

The previous procedure can be generalized in the following way. Let
$D$ be any domain in $\R^d$, which can be partitioned into $m$
congruent subdomains $S_iD$, with $i\in\{0,\cdots,m-1\}$, all of them
{\em similar} to $D$. In our current example, $D=[0,1]\times [0,1]$,
the unit square, which is divided into $m=4$ smaller squares, which we
denote by $S_0D$ (upper-left), $S_1D$ (upper-right), $S_2D$
(lower-left) and $S_3D$ (lower-right).

The action of operators $S_i$ can be iterated. Thus, $S_1S_3D$ denotes
the upper-right quadrant of the lower-right quadrant of the original
square. We can define a {\em geometrical index} as a sequence of
integers $I_G\equiv \{i_k\}_{k=1}^n$, with $i_k\in \{0\cdots
m-1\}$. Each geometrical index denotes a (small) domain $S_{i_1}\cdots
S_{i_n} D$, similar to the original one. In our example, a tiny
square. We can, thus, define a mapping ${\cal S}$ which converts
geometrical indices into regions of $\R^d$ which are similar to $D$:
${\cal S}(I_G)\equiv S_{i_1}\cdots S_{i_n}D$.

Now let us focus on the tensor-product structure of the quantum
Hilbert space. Each state is characterized by a {\em quantum index},
i.e.: a set of $N$ indices taken from certain discrete finite set:
$I_Q\in\Sigma^N$. In our case, $\Sigma=\{0,1\}$. In the case of spin-1
systems, $\Sigma=\{-1,0,1\}$ or, more simply, $\Sigma=\{-,0,+\}$.

The last piece of the scheme is a function ${\cal M}$ mapping quantum
into geometrical indices, ${\cal M}:\Sigma^N \mapsto
\{0,\cdots,m-1\}^n$, such that $I_G={\cal M}(I_Q)$. In our case, this
function groups the quantum indices in pairs, and combines each pair
into a single geometrical index with the simple binary mapping: $00\to
0$, $01\to 1$, $10\to 2$, $11\to 3$. It should be noted that
$n=N/2$. This mapping should be bijective, so as not to lose
information.

Now, the full wavefunction plotting scheme ${\cal K}$ is defined by
providing the original region, $D$, the set of similarity
transformations, $\{S_i\}$, $i\in\{0,\cdots, m-1\}$ and the indices
mapping function ${\cal M}$. Thus, ${\cal K}(I_Q)$ will denote the
region in $\R^d$ obtained by applying ${\cal S}$ to the geometrical
index associated to $I_Q$, i.e.: ${\cal K}(I_Q)={\cal S}({\cal
  M}(I_Q))$. Those cells make up a partition of $D$. It is easy to
prove the essential properties:

\begin{eqnarray}
\cup_{Q\in\Sigma^N} {\cal K}(I_Q) =D \nonumber \\ 
{\cal K}(I_Q) \cap {\cal K}(I_Q') = \emptyset 
\quad\Longleftrightarrow\quad I_Q\neq I_Q'
\label{mapping.properties}
\end{eqnarray}

Thus, for every $x\in D$, there exists a single $I_Q\in\Sigma^N$ such
that $x\in {\cal K}(I_Q)$. This property ensures that we have can
pull-back wavefunctions, i.e.: functions $\psi:\Sigma^N\mapsto \C$,
into complex-valued functions on $D$, ${\cal K}(\psi):D\mapsto\C$.

So, can we devise other possible plotting schemes? Will they make
different properties apparent? We will approach those questions in the
rest of this section.

\subsection{1D plot}

The simplest possible plotting scheme can be realized in 1D for
qubits. Let $D$ be the $[0,1]$ segment, split every time into two
halves: $S_0$ selects the left part, and $S_1$ the right one. Now, the
resulting ${\cal K}$ mapping is equivalent to a binary {\em
  lexicographical} ordering of the wavefunction components. More
explicitly: divide the domain $[0,1]$ into $2^n$ equal cells, index
them from $0$ to $2^n-1$ and attach to each of them the wavefunction
component with the same associated index. 

\begin{figure}
\begin{center}
\epsfig{file=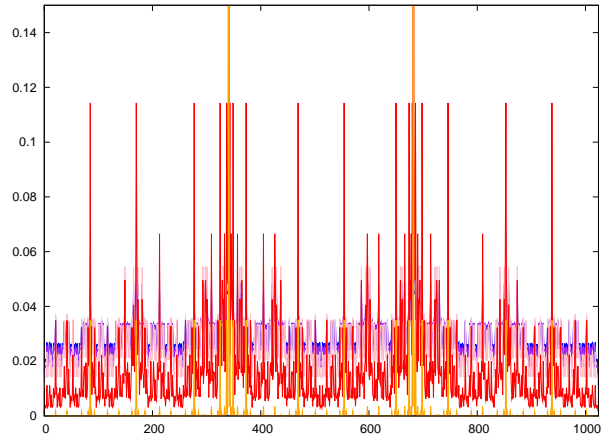,width=6cm,angle=270}
\end{center}
\caption{\label{fig.itf1d}1D representation of the AF ITF ground state
  with $N=10$ qubits and PBC for a few values of $\Gamma$.}
\end{figure}

Figure \ref{fig.itf1d} shows plots (1D) of the ground state of the
antiferromagnetic ITF model, equation \ref{itf.model}, for several
values of $\Gamma$. This plotting scheme is, evidently, much less
appealing than the bidimensional ones. On the other hand, its
simplicity is helpful when attempting to clarify some of the features,
e.g. the Fourier analysis made in section \ref{selfsim}.

\subsection{Spin-1 plots: AKLT states}

As we have already stated, the quantum indices can be built upon any
local set of quantum numbers. For a set of spin-1 particles, the
choice is $\Sigma=\{-1,0,1\}$ or, more simply, $\Sigma=\{-,0,+\}$. If we
start with the same domain, $D=[0,1]^2$, the natural decomposition is
into $3\times 3$ subdomains, as shown in the following (cartesian
product) scheme:

\begin{equation}
\begin{array}{|c|c|c|}
\hline
--  & -0 & ++ \\ \hline
0-  & 00 & 0+ \\ \hline
+-  & +0 & ++ \\ \hline
\end{array}
\label{mapping.spin1}
\end{equation}

Of course, this is not the only possible mapping. With this one, we
have chosen to show the structure of the Affleck-Kenedy-Lieb-Tasaki
(AKLT) state \cite{aklt.88}. It is the ground state of the following
Hamiltonian:

\begin{equation}
H=\sum_{i=1}^N {\vec S}_i\cdot {\vec S}_{i+1} + 
{1\over 3} ({\vec S}_i\cdot {\vec S}_{i+1})^2
\label{aklt.ham}
\end{equation}

This state is an example of {\em valence bond solid} (VBS), and has
attracted considerable attention because of its relation to the
Haldane conjecture \cite{haldane.83}, its non-local order parameter
\cite{dennijs.89} and as a source of inspiration of tensor-network
states \cite{perezgarcia.08}.

\begin{figure}
\begin{center}
\epsfig{file=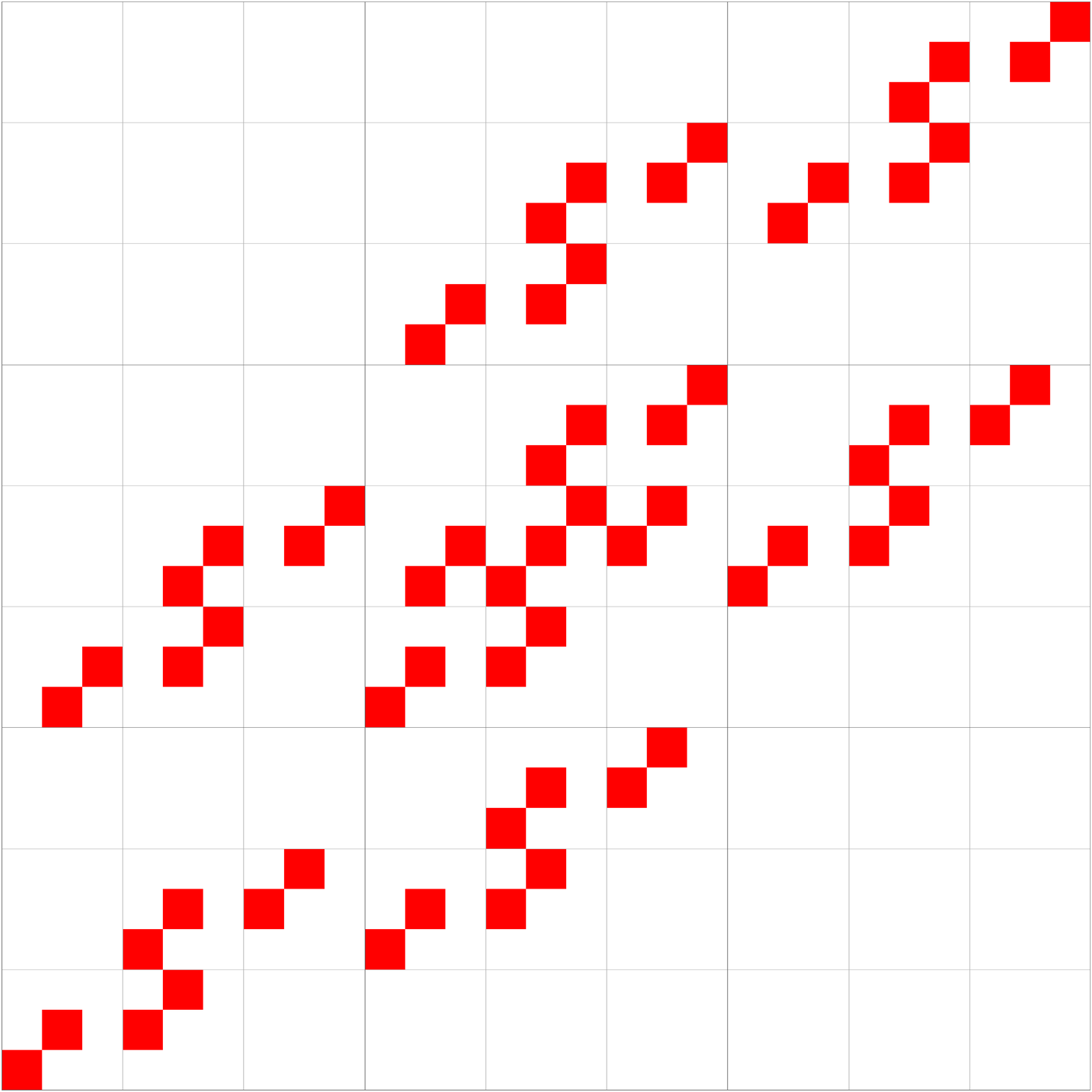,width=4.6cm}
\epsfig{file=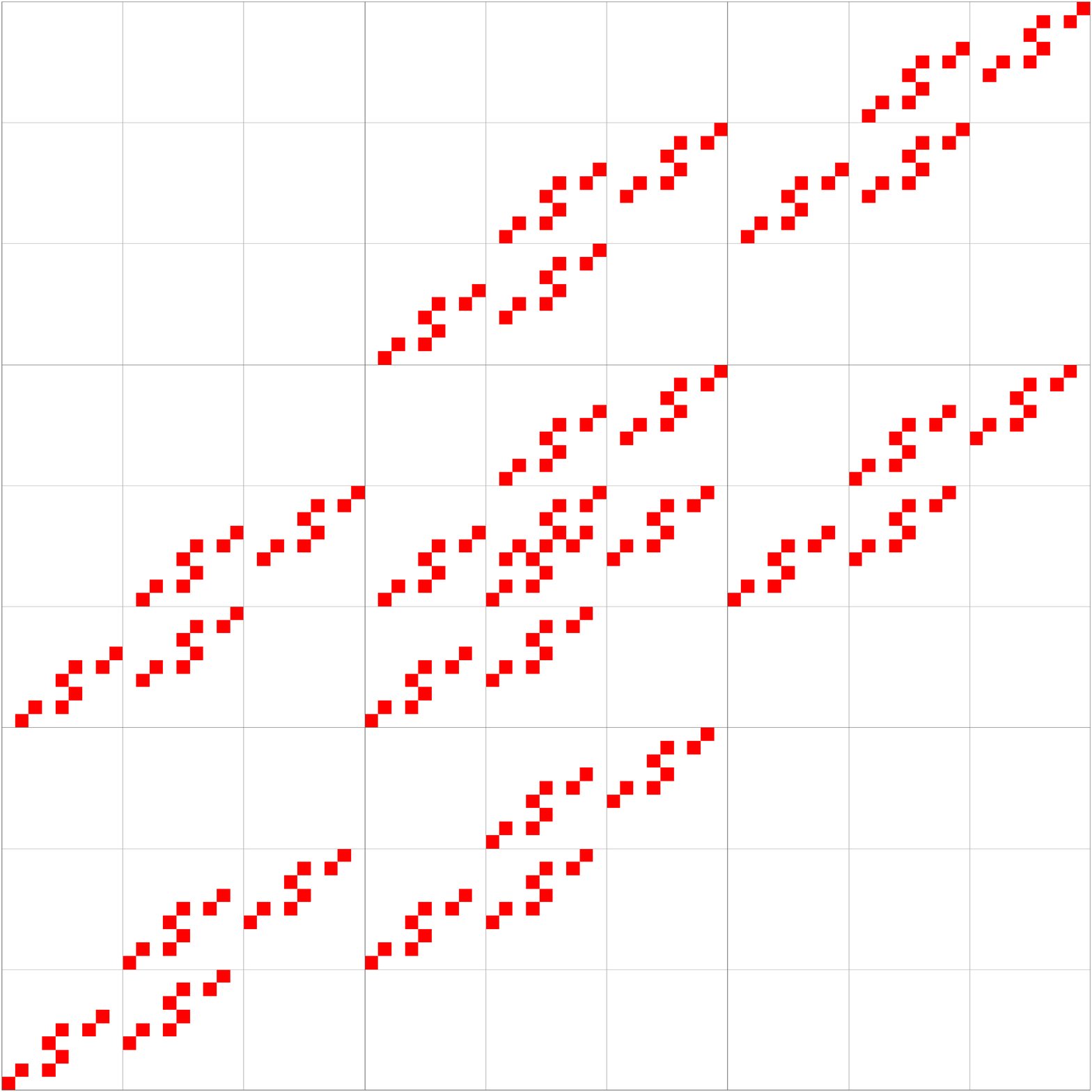,width=4.6cm}
\epsfig{file=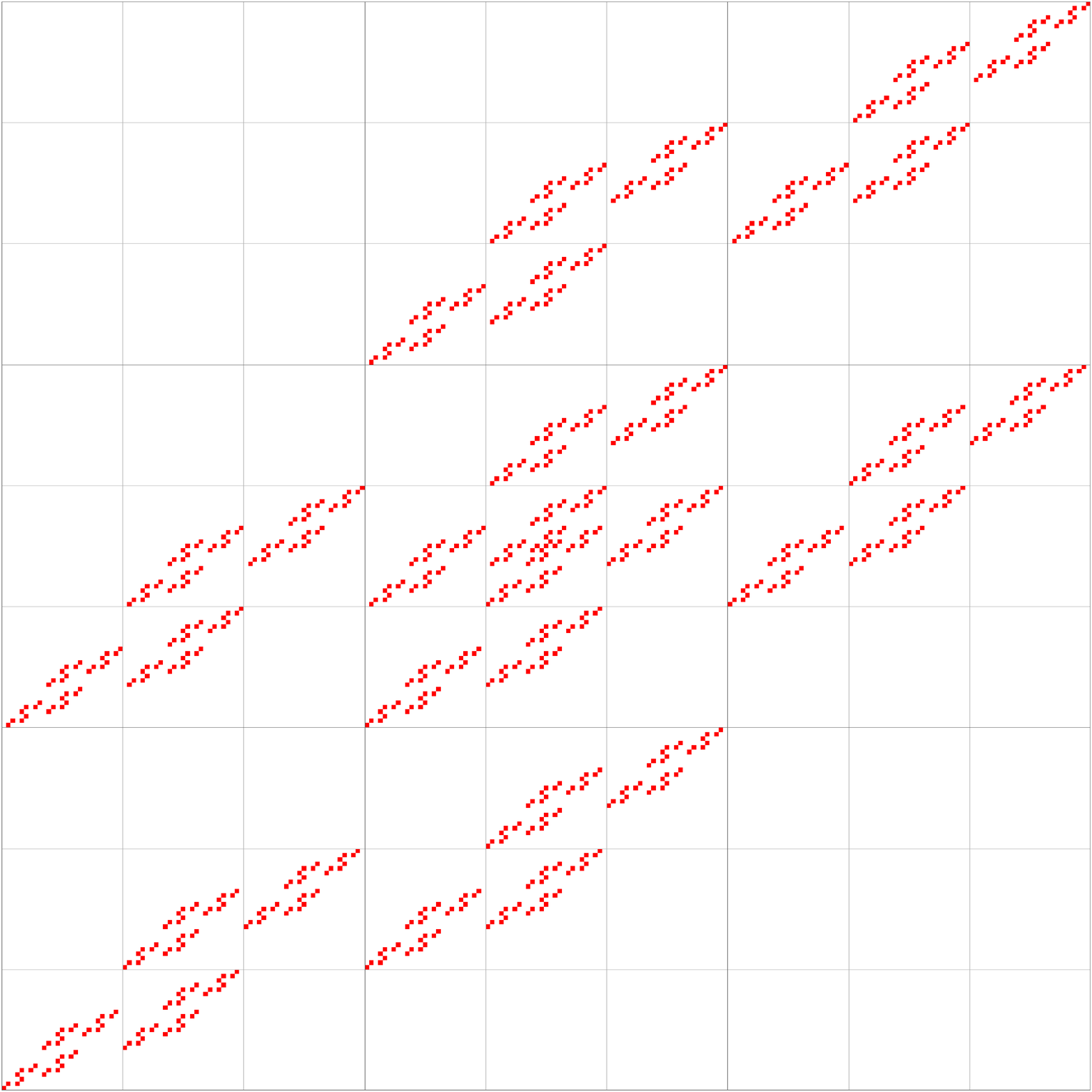,width=6cm}
\end{center}
\caption{\label{fig.aklt} Ground state of the AKLT spin-1 Hamiltonian,
  for $N=6$, $8$ (top) and $10$ (bottom) spins. Notice how the fractal
  structure develops.}
\end{figure}

The result can be seen in figure \ref{fig.aklt} where, for better
visualization, we have marked only the non-zero components of the
wavefunction. Notice the strong self-similarity appearance.  

\subsection{Alternative Square Plot}

Restricting ourselves to qubits and ${\cal D}=[0,1]^2$, it is still
possible to have another inequivalent plotting scheme, by changing the
assignments:

\begin{equation}
\begin{array}{ccc}
00 \to \hbox{Upper left} & 01 \to  \hbox{Upper right} \\
11 \to \hbox{Lower left} & 10 \to  \hbox{Lower right} \\
\end{array}
\label{mapping2}
\end{equation}

In this new plotting scheme the two left corners (top and bottom)
represent the FM states, and the right corners the N\'eel states.

It can be shown that these two are the only possible inequivalent
plotting schemes for qubits on $[0,1]^2$. The reason is the
following. There are $4!=24$ possible associations between
$\{00,01,10,11\}$ and the four quadrants. The group of symmetries
contains three rotations, two reflections on the horizontal and
vertical axes and two reflections on the two diagonals, i.e.: 12
different elements. This leaves only $4!/12=2$ inequivalent choices.

\begin{figure}
\begin{center}
\epsfig{file=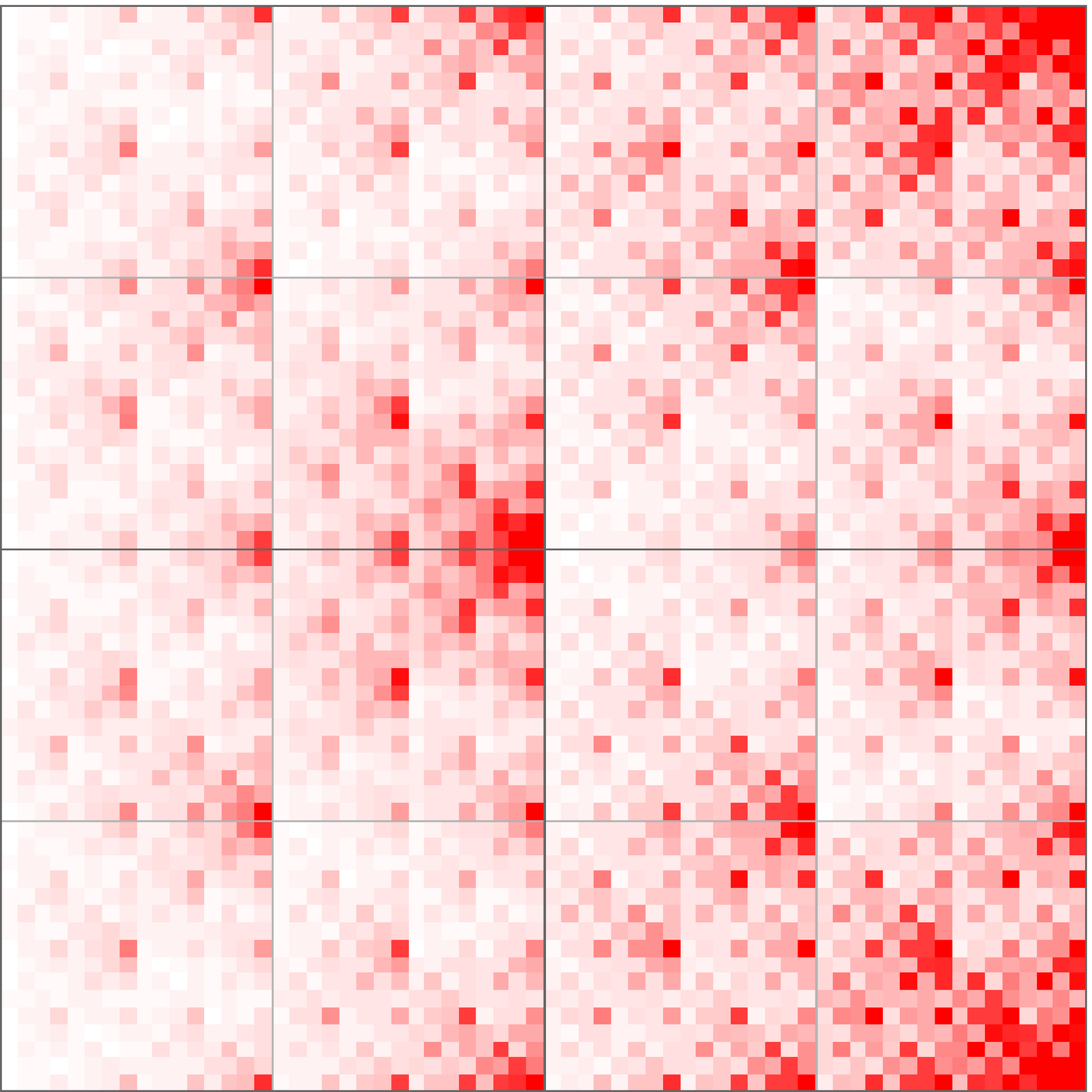,width=6cm}
\epsfig{file=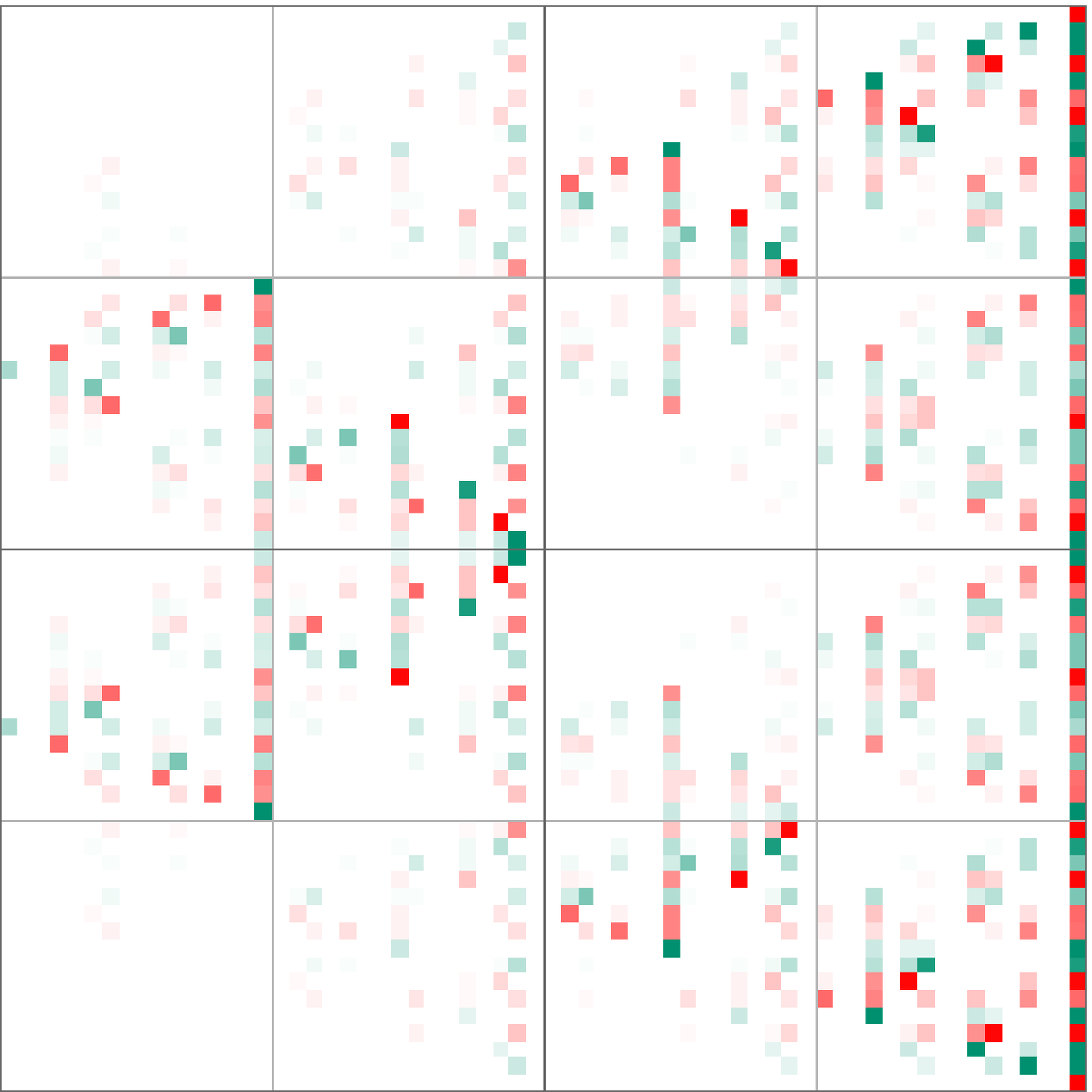,width=6cm}
\end{center}
\caption{\label{fig.alt}Alternative 2D representations, following
  equation \ref{mapping2}, with $N=12$ qubits and PBC. Left: Ground
  state of the critical AF ITF model. Right: Ground state of the
  Heisenberg model. Compare, respectively, to figures \ref{fig.itf}
  (central) and \ref{fig.heis}.}
\end{figure}

As an example, figure \ref{fig.alt} (left) shows the ground state of
the critical ($\Gamma=1$) ITF model with $N=12$ qubits
(eq. \ref{itf.model}). It is therefore, an alternative pictorial
representation of figure \ref{fig.itf}. Figure \ref{fig.alt} (right)
shows the ground state of the Heisenberg model with $N=12$ qubits, in
the new plotting scheme. The N\'eel states are now situated in the
lower and upper right corners. Therefore, the main diagonal line,
hallmark of the spin-liquid structure, lies now in the rightmost
vertical line. The secondary diagonals, on the other hand, are now
dispersed, in a Sierpi\'nski-like structure.

It is apparent that figure \ref{fig.alt} (bottom) is {\em smoother}
than its counterpart, figure \ref{fig.itf} (central). All possible
plotting schemes are equally valid, in principle, just as a polar and
a cartesian representation of the same function are. Can we provide
some sense of {\em plotting quality}? Perhaps: a smoother plot
suggests that the neighbourhood structure of the original wavefunction
is respected more properly by the plotting scheme.

\subsection{Triangular Scheme}

\begin{figure}
\begin{center}
\epsfig{file=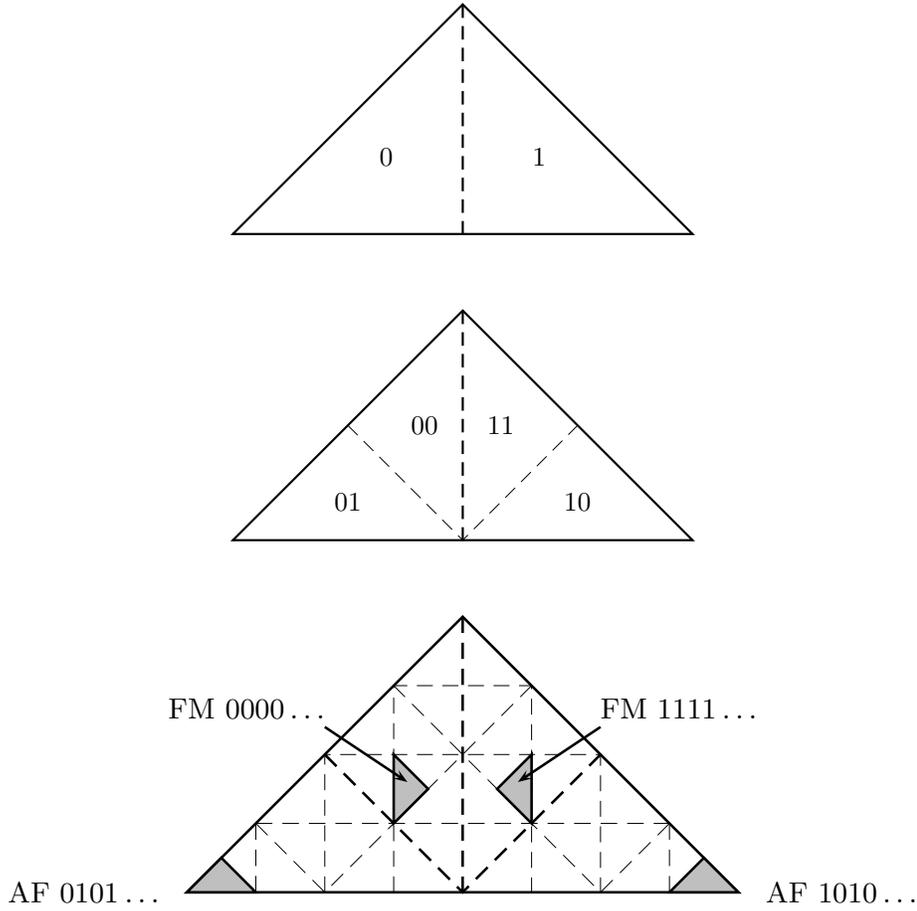,height=120mm}
\end{center}
\caption{\label{fig.triangle} Plotting scheme using rectangular
  triangles. Top and center: first two iterations. Bottom: First four
  iterations (dashed lines). The two N\'eel states of the fifth
  iteration are the bottom left and right filled corner triangles. The
  two FM states correspond to the symmetrically placed filled
  triangles near the center.}
\end{figure}

It is possible to design a 2D plotting scheme of qubits which does not
require grouping the quantum indices in pairs. Let $D$ be a
rectangular isosceles triangle of unit side, with vertices at
$(-1,0)$, $(0,1)$ and $(1,0)$. It can be split into two similar
triangles, of side $1/\sqrt{2}$. Let $S_0$ and $S_1$ be the operators
which select the left and right triangles (as seen when the
right-angle vertex is up). Figure \ref{fig.triangle} shows how such a
representation maps bits into cells. Within this scheme the N\'eel
states go towards the left and right bottom corners. The FM states
correspond to two points near the center, symmetrically placed with
respect to the height. In the thermodynamical limit, these FM points
can be obtained summing a geometrical series: $(\pm 1/5, 2/5)$.

\begin{figure}
\begin{center}
\epsfig{file=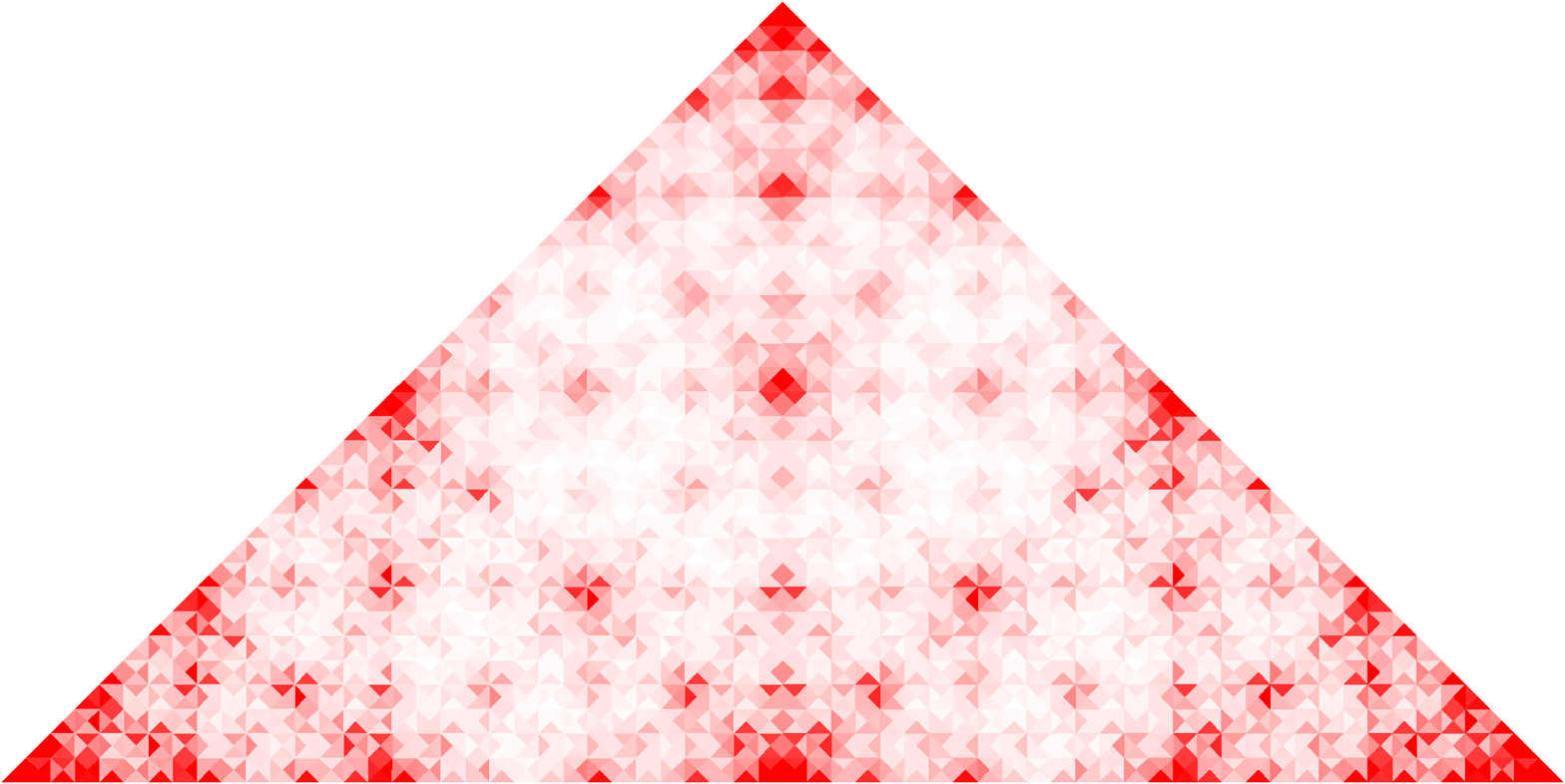,width=8cm}

\vspace{5mm}

\epsfig{file=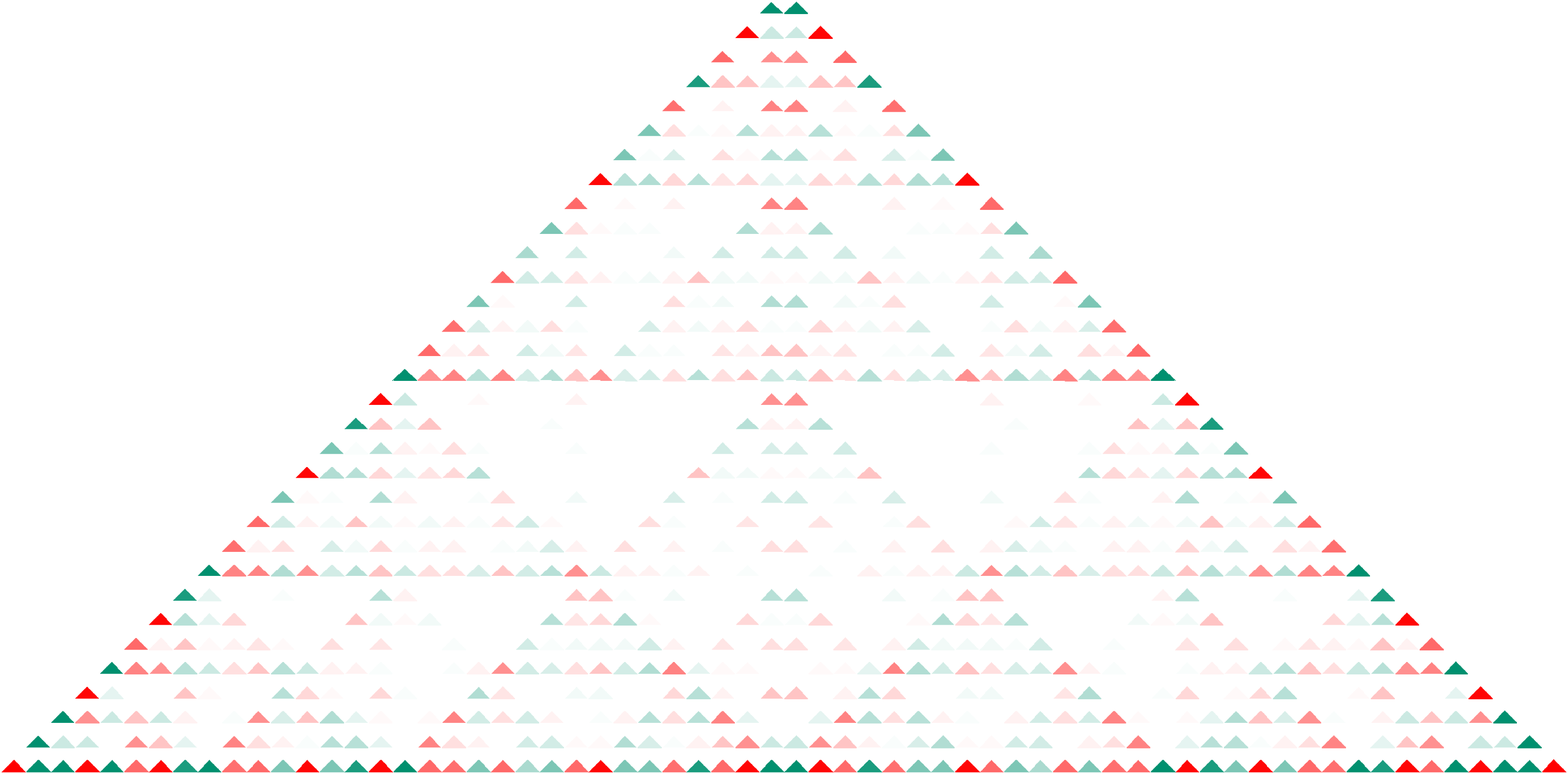,width=8cm}

\vspace{5mm}

\epsfig{file=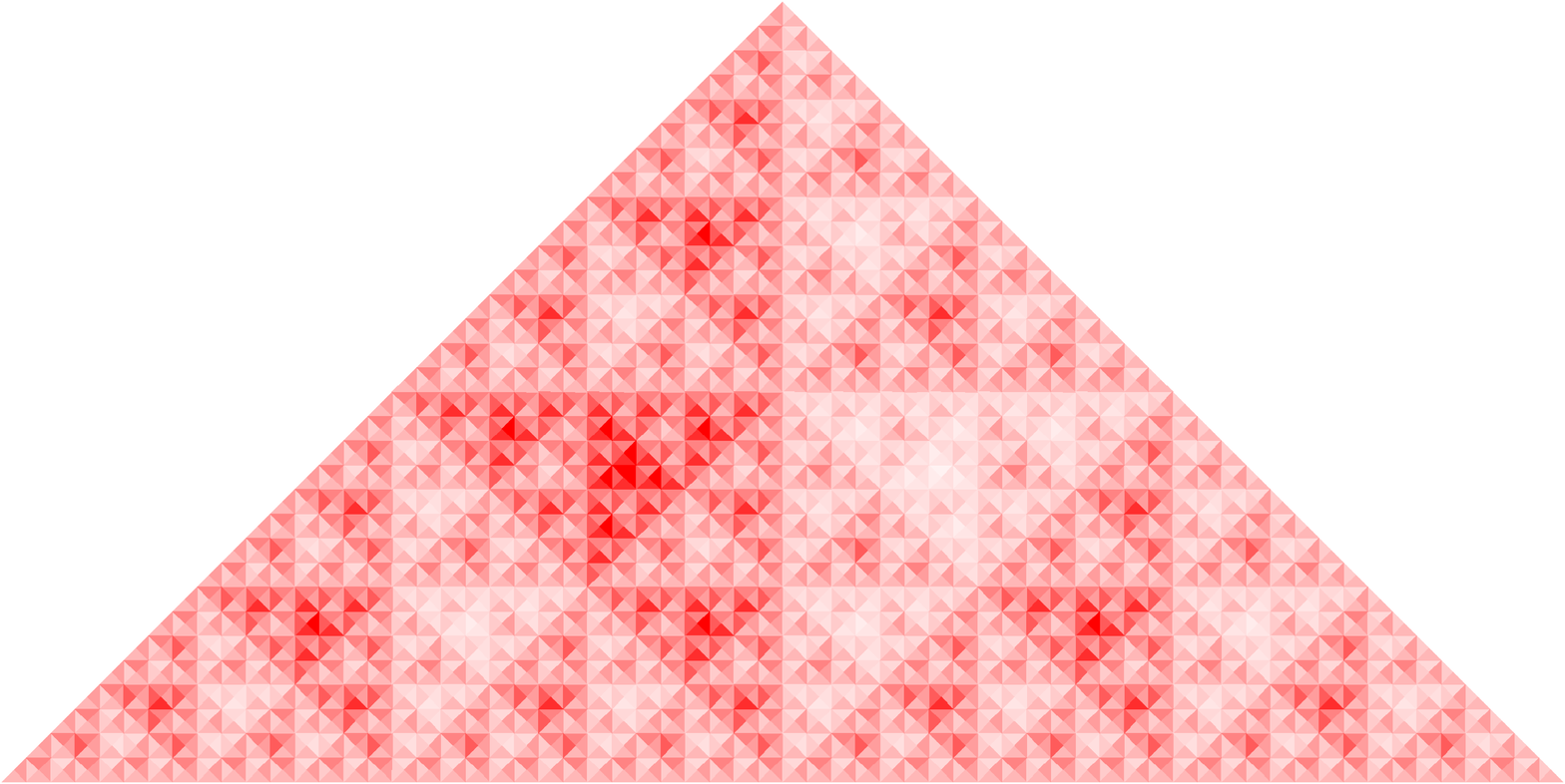,width=8cm}
\end{center}
\caption{\label{fig.wftri}Triangular representations of many-body
  wavefunctions. Top: GS of the AF-ITF model with $N=12$ qubits and
  PBC. Center: GS of the Heisenberg model with $N=12$ qubits and
  PBC. Bottom: product state with $N=12$.}
\end{figure}

Figure \ref{fig.wftri} depicts the ground states of the critical ITF
and the Heisenberg model, and a product state. Notice that the
diagonal lines in the original representation for the Heisenberg GS
have mapped now to the {\em perimeter} of the triangle. The main
diagonal is the hypotenuse, the two secondary diagonals are the other two
sides. The remaining structure, which was not quite clean in the
original representation, comes here as a Sierpi\'nski-like structure. 

\subsection{More Exotic Plotting Schemes}

We will now propose other plotting schemes in order to show the
versatility of the procedure. 

In the case of spin-1 systems, the only alternative that we have found
in order to make the quantum and the geometrical indices coincide is
to work on a Sierpi\'nski triangle. The original domain is, this way,
naturally split into three similar domains: $S_-D$, $S_0D$ and
$S_+D$. Nonetheless, it has the disadvantage that the domain is not
simply connected.

Even more exotic plotting schemes are conceivable. Let $A_0$ be a
regular hexagon. Now proceed to build $A_1$ as the union of $A_0$ and
six congruent hexagons built upon its sides. Repeating the scheme, and
rescaling at each step, we reach a fixed point: $A_\infty$, with the
following property: it can be split naturally into $7$ similar cells
of exactly the same shape \cite{schroeder.92}.


\section{\label{selfsim}Self-similarity of the wavefunction plots}

The plotting schemes described in the present article are evidently
self-similar. It is obvious that the first qubit determines the
largest-scale properties of the plot, and subsequent qubits determine
lower scales properties. The question that we will address is: how
does this self-similarity of the scheme map into fractal or
self-similar properties of the wavefunction plots?

\subsection{Translation-invariance and self-similarity}

Let ${\cal R}$ be the cyclic right-translation shift. A wavefunction
has translational symmetry if, for any quantum index $I_Q$,
$|\psi(I_Q)|=|\psi({\cal R}(I_Q))|$. Does this symmetry bear any visual
consequences in the wavefunction plots? 

In a translationally invariant system, a measurement performed on the
first two qubits and another performed on the last two should have the
same effects. Let us focus on a given possible outcome of the
measurement, e.g.: $00$. Now, the wavefunctions describing the rest of
the system should coincide. If the measured qubits are the first two,
the new wavefunction-plot is obtained by selecting the upper-left
quadrant of the original plot. On the other hand, if the measurement
has been done on the last two qubits, we should {\em decimate}: group
the plot pixels into $2\times 2$ blocks, and select the upper-left
pixel out of each block. Both wavefunctions should coincide, as a
result of translation-invariance.

\begin{figure}
\begin{center}
\epsfig{file=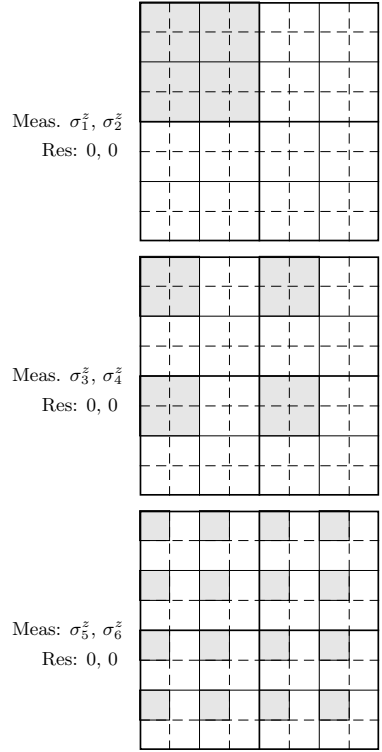,height=10cm}
\end{center}
\caption{\label{fig.translinv}Illustrating translation invariance
  properties in a qubistic plot. All three plots represent the
  qubistic representation of a $N=6$ wavefunction. Assume we measure
  $\sigma^z$ on qubits 1 and 2, and the results are 0 and 0. Then,
  after the measurement, the wavefunction-plot will be given by the
  marked cells of the top plot. The central and bottom plots are
  equivalent, but with measurements performed on qubits $\{3,4\}$ and
  $\{5,6\}$ respectively. If the wavefunction is
  translation-invariant, all three resulting wavefunction-plots should
  be exactly equal.}
\end{figure}

So, plots of translation-invariant wavefunctions display
self-similarity in the following sense. Divide the plot into a matrix
of $2^k\times 2^k$ sub-plots ($k\in\{0,\cdots n-1\}$) and do a further
division of each sub-plot into $2\times 2$ quadrants. Selecting the
same quadrant from each sub-plot and rebuilding a full image will
yield the same result, for all possible values of $k$. Figure
\ref{fig.translinv} illustrates the criterion.

\subsection{Measures of scale invariance}

\begin{figure}
\begin{center}
\epsfig{file=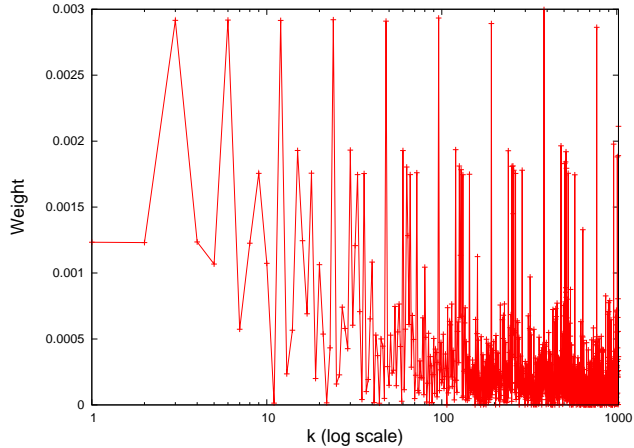,width=6cm,angle=270}
\end{center}
\caption{\label{fig.ft} Fourier transform of the ground state of the
  1D plot of the critical ITF model, with $N=10$ qubits and PBC, for
  $\Gamma=0.6$.}
\end{figure}

Scaling invariance of the wavefunction plots should also be visible in
the Fourier transform. In effect, figure \ref{fig.ft} shows the
transform of a 1D-plot of an AF ITF Hamiltonian (eq. \ref{itf.model})
with $N=10$ qubits and PBC. The momenta are displayed in logarithmic
scale, and we can spot a clear periodic structure. Evidently, exact
log-periodicity is impossible to achieve since each period contains a
larger number of degrees of freedom than the preceding one. This
feature is visible for a wide range of transverse fields, i.e.: it is
not linked to criticality.

Another interesting indicator of self-similarity is provided by the
R\'enyi fractal dimensions \cite{Halsey.86}. Let us consider the
probability distribution associated to a wavefunction plot (taking the
modulus squared), $P_N=\{p_{N,i}\}$ for $N$ qubits. We can compute
the R\'enyi entropy of order $q$, i.e.:

\begin{equation}
R_q(P^{(N)})\equiv {1\over 1-q} \log\( \sum_i p_{N,i}^q \)
\label{renyi.entr}
\end{equation}

\noindent Now we define the R\'enyi dimensions by:

\begin{equation}
d_q \equiv \lim_{N\to\infty} { R_q(P_N) \over \log\(b^{N/2}\) } 
\label{renyi.dim}
\end{equation}

\noindent where $b$ is $2$ for qubits or $3$ for spin-1. With this
notation, $d_0$, $d_1$ and $d_2$ are, respectively, the support,
information and correlation dimensions of the fractal. The full set of
$d_q$ provide the same information as the {\em multifractal spectrum}.

\begin{figure}
\begin{center}
\epsfig{file=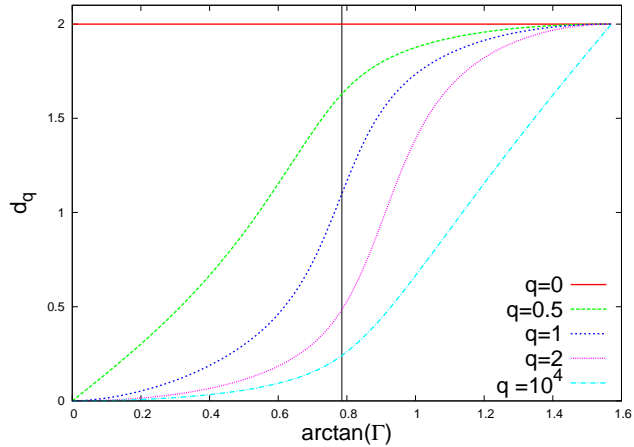,width=6cm,angle=270}
\end{center}
\caption{\label{fig.renyi.itf}R\'enyi fractal dimensions (equation
  \ref{renyi.dim}) of the wavefunction plot of the ground state of the
  AF ITF model with PBC. Notice that the $X$-axis is labeled with
  $\arctan(\Gamma)$ in order to highlight the symmetry. The vertical
  bar shows the critical case, $\Gamma_c=1$. Computations are carried
  out for $N=10$ qubits, but the results are quite independent of the
  system size.}
\end{figure}

Figure \ref{fig.renyi.itf} shows a few R\'enyi dimensions for an AF
ITF model as a function of $\Gamma$. In our case, the support
dimension $d_0$ is always $2$, since all probability values are
non-zero. All the other dimensions interpolate between $0$ (for
$\Gamma\to 0$) and $2$ ($\Gamma\to\infty$). The information dimension,
$d_1$ seems to capture most accurately the physical properties of the
model, since its growth rate is maximal at the critical point.

It is an interesting exercise to prove that, for the AKLT state shown
in figure \ref{fig.aklt}, all R\'enyi dimensions with $q>0$ are equal
to $\log(4)/\log(3)\approx 1.26$.


\section{\label{entanglement}Visualization of entanglement}

One of the most intriguing features of quantum many-body systems is
{\em entanglement}. A system is entangled if measurement on one of its
parts affects the results of subsequent measurements on others, even
if they are well separated. Einstein himself described this phenomenon
as {\em ``spukhafte Fernwirkung''} (spooky action at a distance)
\cite{einstein.born}. It is considered as a {\em resource} for quantum
computation and communication \cite{nielsen.04}, as well as providing
very useful insight regarding quantum phase
transitions \cite{vidal.03}.

\subsection{Visual estimate of entanglement}

Is entanglement visualizable from our wavefunction plots?
Yes. Summarizing the results of this section we may say that
entanglement shows as image complexity. Let us consider all quadrants
of level-$k$ within the plot, normalized. If there are only $p$
different quadrants, then the entanglement entropy is $\leq
\log(p)$. Concretely, if all level-$k$ quadrants are equal, the system
is factorizable.

In section \ref{section.factorizable} we discussed product states,
i.e.: systems without entanglement. Let us recall the conclusions
exposed in that section. If the first two qubits are disentangled from
the rest of the system, measurements made upon them should not have
influence on the rest. Therefore: all four quadrants of the plot are
equal (modulo normalization). If all the qubits are disentangled (at
least by pairs), then the result is extended: if the plot is split
into a $2^k\times 2^k$ matrix of subimages, for all $k$, all the
sub-images are equal (modulo normalization). This result can be
expressed in a more concise way: {\em the plot of a product state is
  trivially self-similar}. Every quadrant, of every size, is the same
as any other, after proper normalization.

What happens if the system is entangled? Let us now consider a generic
wavefunction, $\ket|\Psi>$, and split the system into a left and a
right parts, $L$ and $R$. The left part will correspond to qubits 1 to
$2k$ and the right part to qubits $2k+1$ to $N$, for any $k$. We can
always perform a {\em Schmidt decomposition}:

\begin{equation}
\ket|\Psi>=\sum_{i=1}^m \lambda_i \ket|\psi^L_i>\otimes\ket|\psi^R_i>
\label{schmidt.decomposition}
\end{equation}

\noindent where the orthonormal sets $\{\ket|\psi^L_i>\}$ and
$\{\ket|\psi^R_i>\}$ are called the left and right-states, and
characterize the physics of each part, $\lambda_i$ are called the
Schmidt coefficients and $m$ is the Schmidt rank, which is a measure
of entanglement. If $m=1$, the state is factorizable. A state with
Schmidt rank $m$ can not have entanglement entropy larger than
$\log(m)$.

The left part corresponds to the larger scales, and the right part to
the smaller ones. Let us make this statement concrete.

Consider the Hilbert space for the left part, and let $\{\ket|x>\}$ be
the basis of tensor states for it. E.g.: if $k=1$, the left part has
two qubits and the states $\{\ket|x>\}$ are $\ket|00>$, $\ket|01>$,
$\ket|10>$ and $\ket|11>$. Now we will consider what are their
geometric counterparts in the wavefunction plot. Within the original
2D plotting scheme, qubits 1 to $2k$ correspond to the first $k$
quadrant divisions. Let us divide the original plotting square into a
matrix of $2^k\times 2^k$ quadrants. Each tensor state $\ket|x>$ can
be attached to one of these quadrants, which we will denote simply by
$x$.

The left-states, $\ket|\psi^L_i>$ can be expressed as

\begin{equation}
\ket|\psi^L_i>=\sum_{x} \psi^L_{ix} \ket|x>
\label{left.states}
\end{equation}

Now, let us focus on the right part. Each right-state, $\ket|\psi^R_i>$
can be plotted inside a level-$k$ quadrant using the standard
representation. Let us call the corresponding plot $R_j$.

What is the actual image, for the full wavefunction plot, on the
$x$-th quadrant? Inserting equation \ref{left.states} into the Schmidt
decomposition \ref{schmidt.decomposition} we can see that it is given
by the expression

\begin{equation}
C(x)=\sum_{i=1}^m \lambda_i \psi^L_{ix} R_i
\label{final.image}
\end{equation}

The conclusion is that, {\em for each level-$k$ quadrant the plot is a
  linear combination of the $m$ right-state plots}, with weights given
by the the left-states components and the Schmidt coefficients.

Therefore, level-$k$ quadrants within the final plot are built upon
only $m$ {\em fundamental images}, or building bricks, which are the
plots of the right-states. In other terms: {\em the Schmidt rank $m$
  for a given left-right partition coincides with the effective
  dimension of the subspace spanned by all images in level-$k$
  quadrants}. This statement provides a way to give a coarse estimate
the entanglement of the wavefunction: if, at level-$k$, the number of
{\em different} quadrants is $p$, then the block of the first $2k$
qubits has a Schmidt rank of $m\leq p$, and the entanglement entropy
is $S\leq\log(p)$. As a corollary, if all quadrants are exactly the
same, then $m=1$ and $S=0$, the system is factorizable, as we already
stated.

The logic behind the estimate is to find the number of different
building blocks at each scale. If we want to be precise, the Schmidt
rank is given by the dimension of the subspace spanned by all quadrant
images at a certain level, but this value is much more difficult to
estimate visually.

\begin{figure}
\begin{center}
\begin{tabular}{cccccc}
 &
(A) &
(B) &
(C) &
(D) &
(E) \\
\raisebox{0.65cm}{$\sigma^z$ basis} &
\epsfig{file=blocks_sep_01.eps,width=1.5cm} &
\epsfig{file=blocks_ghz_01.eps,width=1.5cm} &
\epsfig{file=blocks_w_01.eps,width=1.5cm} &
\epsfig{file=blocks_d2_01.eps,width=1.5cm} &
\epsfig{file=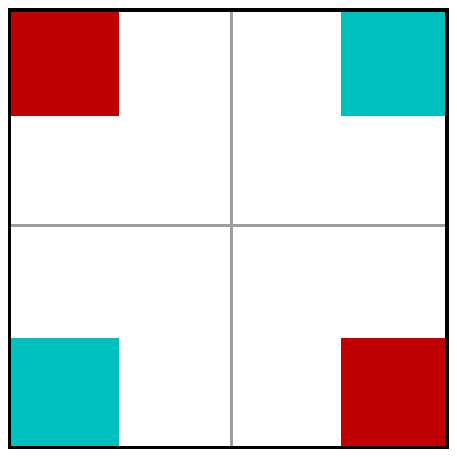,width=1.5cm}\\
\raisebox{0.65cm}{$\sigma^x$ basis} &
\epsfig{file=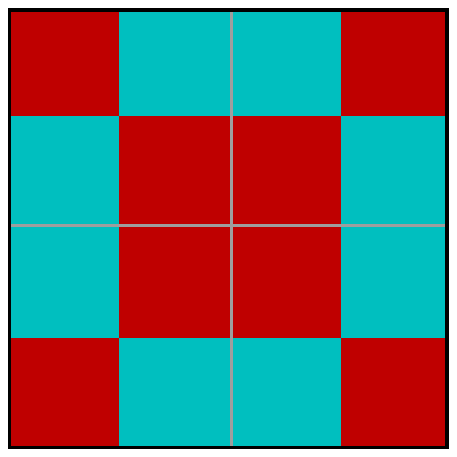,width=1.5cm} &
\epsfig{file=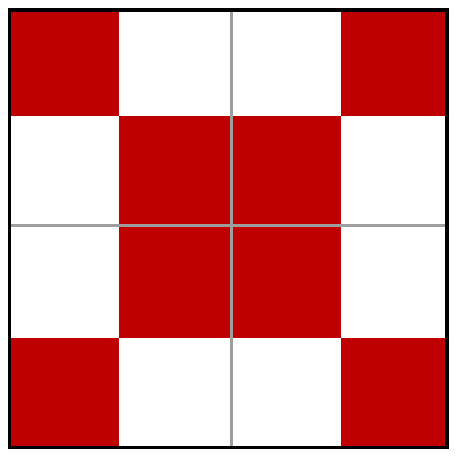,width=1.5cm} &
\epsfig{file=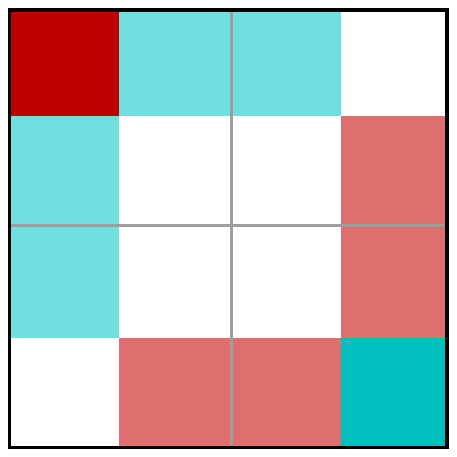,width=1.5cm} &
\epsfig{file=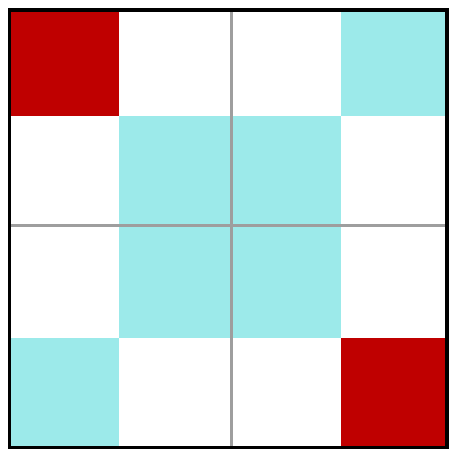,width=1.5cm} &
\epsfig{file=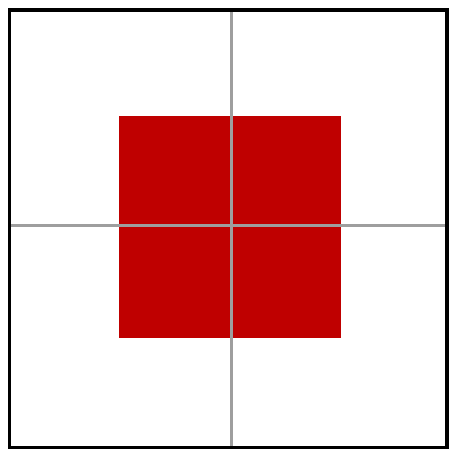,width=1.5cm}\\
Schmidt rank: &
1 &
2 &
2 &
3 &
4 \\
Von Neumann entropy: &
$0$ &
$\log 2 = 1$  &
$\log 2 = 1$  &
$\log 3-\frac{1}{3}$  &
$\log 4 = 2$ \\
&
&
&
&
$\approx 1.25$ 
\end{tabular}
\end{center}
\caption{\label{fig.blockent} Visual estimates of entanglement using
  qubistic plots for some relevant states of $N=4$ qubits. The first
  row shows the usual qubistic plots in the $\sigma^z$ basis. The
  second row, on the other hand, depicts the plots using the
  $\sigma^x$ basis. Since entanglement is invariant under local
  changes of basis, the visual estimate of entanglement should not
  change. The third row provides the Schmidt rank in all cases for the
  separation between the first and second pairs of qubits, and the
  fourth row contains the von Neumann entropy (logarithms are always with
  basis 2). The states correspond to the columns: A.- $\ket|0000>$;
  B.- GHZ state; C.- W state; D.- Dicke state at half filling; E.-
  $\ket|0000>+\ket|1111>-\ket|1010>-\ket|0101>$.}
\end{figure}

Let us apply the estimate in a set of simple cases, with $N=4$ qubits
(i.e.: only two levels). Figure \ref{fig.blockent} shows the qubistic
plots for a set of states similar to those of figure
\ref{fig.n4}. Since entanglement is invariant under local changes of
basis, we also show the qubistic plot in the basis of eigenstates of
$\sigma^x$. Both plots provide a similar estimate, which is compared
in each case with the exact value. (A) The state $\ket|0000>$ is
factorizable, which can be seen in both plots. In the $\sigma^z$ plot,
only one of the sub-images is non-zero. In the $\sigma^x$ picture, all
four sub-images are the same, modulo a sign. (B) The GHZ is not
factorizable. In both basis can be seen that the number of different
sub-images is 2. (C) corresponds to the W state, which is a bit more
complex. In the $\sigma^z$ basis it is evident that the number of
different sub-images is 2, which corresponds to the Schmidt rank. In
the $\sigma^x$ basis, the visual estimate gives 3 different
sub-images. Our prediction is still valid, since the estimate only
provides an upper bound. The reason for the error is that the 3
sub-images are not {\em linearly independent}. This example serve as a
warning: some basis may provide clearer visual estimates than
others. (D) Is the Dicke state at half-filling. In this case the
visual estimate coincides for both basis, 3 different sub-images. But
do not have the same {\em weight}, and the von Neumann entropy is
smaller than $\log(3)$. (E) The
$\ket|0000>+\ket|1111>-\ket|1010>-\ket|0101>$ state has four different
sub-images in both basis, and achieves maximal Schmidt rank and
entanglement entropy.

\begin{figure}
\begin{center}
\begin{tabular}{cc}
\epsfig{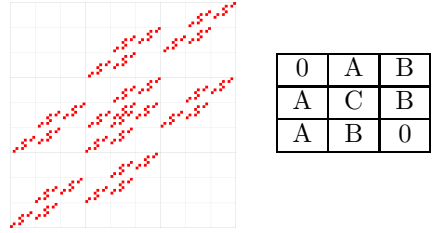} &
\raisebox{1cm}{
\begin{tabular}[b]{|c|c|c|}
\hline
0 & A & B \\ \hline
A & C & B \\ \hline
A & B & 0 \\
\hline
\end{tabular}}
\end{tabular}
\end{center}
\caption{\label{fig.aklt.entang}AKLT qubistic plot for $N=8$ spins
  and the associated sub-image pattern: boxes with the same letter
  contain (very approximately) the same sub-image.}
\end{figure}

The strategy can be applied to the AKLT state, depicted in figure
\ref{fig.aklt}. At any splitting level the {\em exact} number of
different images is always 5. But, as the number of sites increases,
some of these images become more and more alike, until only 3 of them
are distinguishable. Figure \ref{fig.aklt.entang} shows the sub-image
pattern more clearly. See, for example, the $-+$ and $0+$ quadrants of
the plots in figure \ref{fig.aklt}: their differences are easy to spot
for $N=6$, but almost unnoticed for $N=10$. This implies that the
Schmidt rank is always $\leq 3$, providing the estimate $S\leq
\log(3)$, independent of the depth level, which is exactly the actual
value in the thermodynamic limit \cite{fan.04}.

\begin{figure}
\begin{center}
\epsfig{file=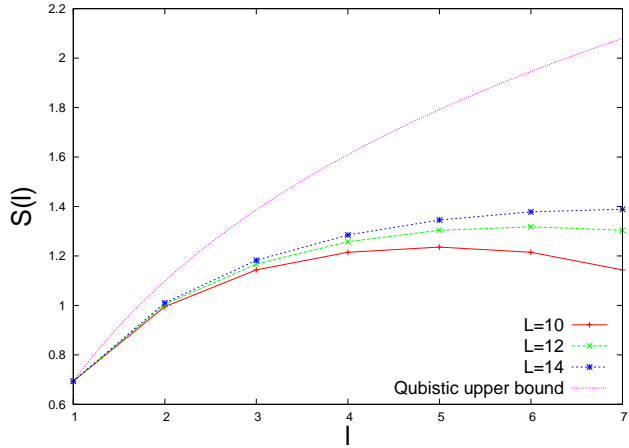,width=6cm,angle=270}
\end{center}
\caption{\label{fig.entropy.dicke}Entanglement entropy of a block of
  $l$ sites in a half-filling Dicke state of $L=10$, $12$ and $14$
  sites, compared to the upper bound obtained from the qubistic plot,
  which is $S(l)\leq\log(l+1)$.}
\end{figure}

On the other hand, taking the half-filling Dicke states of figure
\ref{fig.dicke}, it is evident that, at every magnification level, the
number of different subimages increases by two. Thus, $S(k)\leq
\log(2k+1)$ in terms of levels, or $S(l)\leq \log(l+1)$ for qubits, if
$l\leq N/2$. This bound is found to be fulfilled by the numerical
calculations shown in figure \ref{fig.entropy.dicke}.

The reason for the difference between the estimate and the actual
values of entanglement in fig. \ref{fig.entropy.dicke} is
twofold. First, the number of different level-$l$ quadrants is, in
general terms, a very poor way to estimate the dimension of the
subspace spanned by them. Second, the value estimated this way is just
the Schmidt rank, whose logarithm is just an upper bound to the actual
entanglement entropy. Both problems can be handled within the more
comprehensive framework, described in the next section.

\subsection{Entanglement and the cross-correlation matrix}

Given a wavefunction plot and a level $k$, let us divide the full
region into a grid of $2^k\times 2^k$ sub-plot. Moreover, let $x$ be
an index running through all such sub-plots and $C(x)$ be the actual
image displayed in it, as in equation \ref{final.image}. Now we define
a {\em cross-correlation matrix} for the plot image, $R(x,x')$, as

\begin{equation}
R(x,x')=\langle C(x) | C(x') \rangle
\label{cross-correlation}
\end{equation}

This cross-correlation matrix bears full information about
entanglement of the first $2k$ qubits within the wavefunction, as we
proceed now to show.

According to equation \ref{final.image}, the image on quadrant $x$ is
given by a linear combination of the right-states. Using the
orthogonality property assumed for them we obtain

\begin{eqnarray}
R(x,x')= & \sum_{i,j=1}^m \bra<\psi^R_i|  \psi^{L*}_{ix} \lambda_i
\lambda_j \psi^L_{jx'} \ket|\psi^R_j> \nonumber \\
=  & \sum_{i=1}^m \lambda_i^2 \psi^{L*}_{ix} \psi^L_{ix'}
\label{density.matrix}
\end{eqnarray}

\noindent Thus, we recognize that $R(x,x')$ is just the {\em density
  matrix for the left part}. In other terms:

\begin{equation}
R(x,x')=\rho^L_{xx'}
\label{cc-dm}
\end{equation}

Therefore, {\em the cross-correlation matrix of the wavefunction plot
  holds full information related to entanglement}. 

For example, for a product state, all subimages are equivalent modulo
normalization. Thus, we can assume that

\begin{equation}
R_F(x,x')=N(x)\cdot N(x')
\label{factorizable-cc}
\end{equation}

\noindent with $N(x)=\langle C(x) | C(x) \rangle^{1/2}$ is the norm
for each subimage. Obviously, $\sum_x N^2(x)=1$ and, thus, the matrix
$R_F$ is just a projector on a line. Its spectrum is, in decreasing
order, $\sigma(R_F)=\{1,0,\cdots,0\}$. Therefore, its entanglement
entropy is zero.


\section{\label{pauli}Frame representation}

In this last section we describe a rather different approach to the
problem of providing a graphical representation of a quantum many-body
system, but still self-similar by design. Instead of plotting
wavefunction amplitudes, or probabilities, we can plot the expectation
values of a bidimensional array of operators, chosen in such a way
that the full information contained in the wavefunction is
preserved. This is called a {\em frame representation} of the quantum
state \cite{Ferrie.08}. According to Wootters and coworkers
\cite{Wootters.04}, the final representation may correspond to a
discrete analogue of a {\em Wigner function} \cite{Ferrie.11}, with
very interesting properties in order to characterize {\em
  non-classicality}, such as its {\em negativity} \cite{Spekkens.08}.

Let us consider a system of $n$ qubits, described by a certain density
matrix $\rho$. Now, let us consider the unit square $[0,1] \times
[0,1]$ and any two numbers, $x$ and $y$, characterized by their binary
expansion: $x=0.X_1X_2\cdots X_n$, $y=0.Y_1Y_2\cdots Y_n$. The value
attached to the point $(x,y)$ in the plot will be given by the
expectation value in $\rho$ of the operator $A(x,y)$:

\begin{equation}
f(x,y)=\tr \left[ \rho A(x,y) \right]
\label{traceA}
\end{equation}

\noindent where $A(x,y)$ is given by:

\begin{eqnarray}
A(x,y) &= 
\bigotimes_{k=1}^{n} (-i)^{X_k Y_k} 
\(\sigma_k^x\)^{X_k} \(\sigma_k^y\)^{Y_k}\\
&= \bigotimes_{k=1}^{n} \sigma_k^{X_k+2Y_k}.
\label{defpauliplot}
\end{eqnarray}

In other words, we plot the expected value of every combination of
tensor products of $\{ \sigma^0\equiv
I,\ \sigma^1\equiv\sigma^x,\ \sigma^2\equiv\sigma^y,\ \sigma^3\equiv\sigma^z\}$.
In particular, on the $y=0$ line we get uniquely correlations in
$\sigma_x$; on $x=0$, those in $\sigma_y$, and on $x=y$ those
corresponding to $\sigma_z$. Such representation is unique for every
density matrix, and can be reverted as follows:

\begin{equation}
\rho = \sum_{x,y} \frac{1}{2^n} f(x,y) A(x,y).
\label{inverseformula}
\end{equation}

\begin{figure}
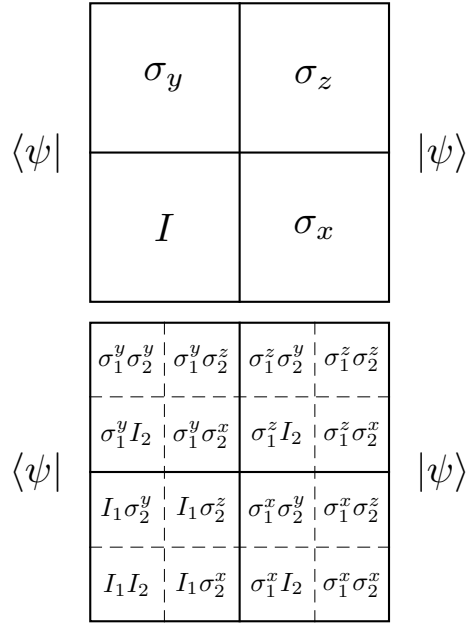

\psset{unit=1mm}
\rput(30,0){%
\rput(-7,-20){\scalebox{1.5}{$\langle\psi|$}}
\rput(47,-20){\scalebox{1.5}{$|\psi\rangle$}}
\psframe(0,0)(40,-40)\psline(0,-20)(40,-20)\psline(20,0)(20,-40)%
\rput(10,-10){\scalebox{1.5}{$\sigma_y$}}
\rput(10,-30){\scalebox{1.5}{$I$}}
\rput(30,-10){\scalebox{1.5}{$\sigma_z$}}
\rput(30,-30){\scalebox{1.5}{$\sigma_x$}}
}
\rput(30,-42.5){%
\rput(-7,-20){\scalebox{1.5}{$\langle\psi|$}}
\rput(47,-20){\scalebox{1.5}{$|\psi\rangle$}}
\psframe(0,0)(40,-40)
\psline(0,-20)(40,-20)\psline(20,0)(20,-40)%
\psline[linestyle=dashed,linewidth=.2pt](10,0)(10,-40)%
\psline[linestyle=dashed,linewidth=.2pt](30,0)(30,-40)%
\psline[linestyle=dashed,linewidth=.2pt](0,-10)(40,-10)%
\psline[linestyle=dashed,linewidth=.2pt](0,-30)(40,-30)%
\rput(5,-5){%
\rput(0,0){$\sigma^y_1\sigma^y_2$}
\rput(10,0){$\sigma^y_1\sigma^z_2$}
\rput(20,0){$\sigma^z_1\sigma^y_2$}
\rput(30,0){$\sigma^z_1\sigma^z_2$}}
\rput(5,-15){\rput(0,0){$\sigma^y_1I_2$}
\rput(10,0){$\sigma^y_1\sigma^x_2$}
\rput(20,0){$\sigma^z_1I_2$}
\rput(30,0){$\sigma^z_1\sigma^x_2$}}
\rput(5,-25){\rput(0,0){$I_1\sigma^y_2$}
\rput(10,0){$I_1\sigma^z_2$}
\rput(20,0){$\sigma^x_1\sigma^y_2$}
\rput(30,0){$\sigma^x_1\sigma^z_2$}}
\rput(5,-35){\rput(0,0){$I_1I_2$}
\rput(10,0){$I_1\sigma^x_2$}
\rput(20,0){$\sigma^x_1I_2$}
\rput(30,0){$\sigma^x_1\sigma^x_2$}}
}
\vspace{85mm}
\caption{\label{fig.frame}Illustrating the frame representation of
  eq. \ref{defpauliplot}. Top: operator assignment for a single
  qubit. Bottom: For two qubits. Products must be understood as tensor
products, with the superscript denoting the qubit index.}
\end{figure}

In order to attain some intuition about the representation, figure
\ref{fig.frame} illustrates it for one and two qubits. At each cell,
we depict the expected value of a ``string'' operator, as shown.

\begin{figure}
\begin{center}
\epsfig{file=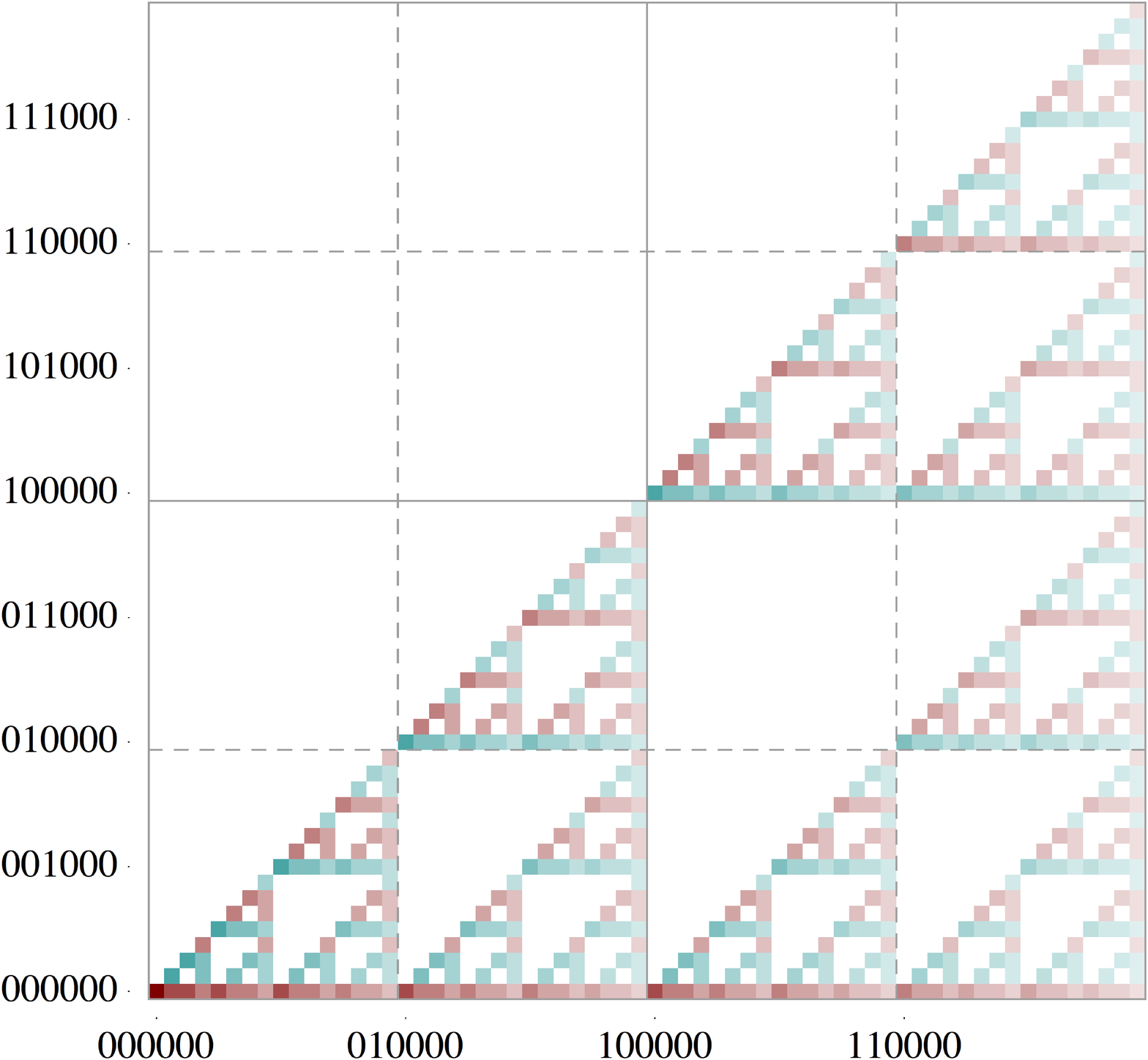,width=6cm}
\end{center}
\caption{\label{fig.pauliprod} Frame representation for a particular
  product state of $N=6$ qubits, described in equation
  \ref{fancyproduct}. Notice the Sierpi\'nski-like structure, which is
  explained in the text.}
\end{figure}

Figure \ref{fig.pauliprod} shows our first example: the frame
representation of a product state given by

\begin{equation}
\ket|\Psi>=\( \cos\({\pi\over 8}\)\ket|0> -
\sin\({\pi\over 8}\)\ket|1> \)^5
\label{fancyproduct}
\end{equation}

\noindent i.e.: a spin pointing half-way between the $-X$ and $Z$
axes. The plot shows a striking Sierpi\'nski-like structure, which can
be fully understood by noticing that, in this state,
$\bra<\Psi|\sigma_x\ket|\Psi>$ and $\bra<\Psi|\sigma_z\ket|\Psi>$ are
nonzero, while $\bra<\Psi|\sigma_y\ket|\Psi>=0$. If, in figure
\ref{fig.frame} (bottom) we cross out all elements with a $\sigma_y$,
the Sierpi\'nski-like structure will appear. Self-similarity,
therefore, is rooted in the plotting scheme, as in the previous case.

\begin{figure}
\begin{center}
\begin{tabular}{c}
\epsfig{file=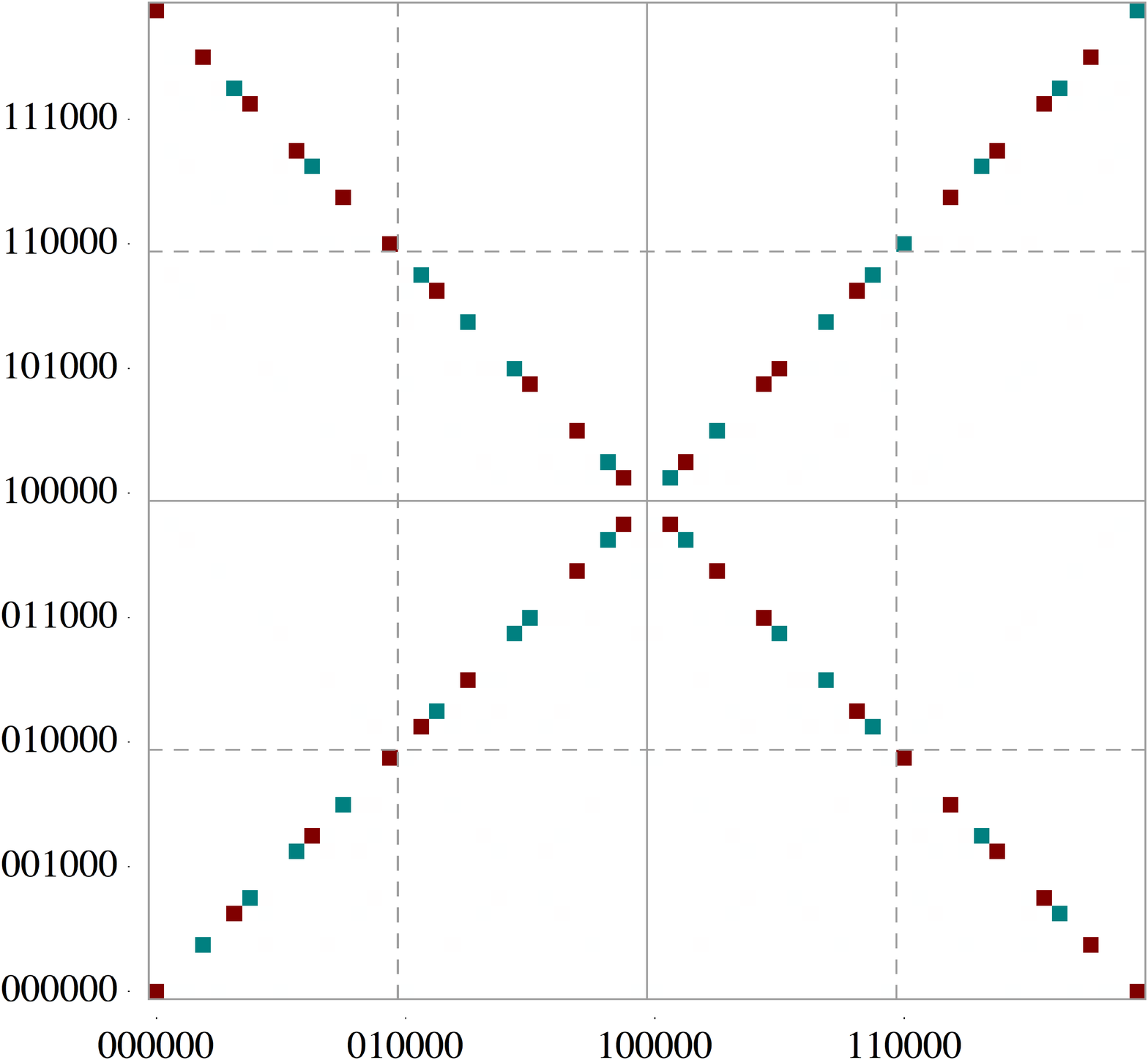,width=5cm} \\
\epsfig{file=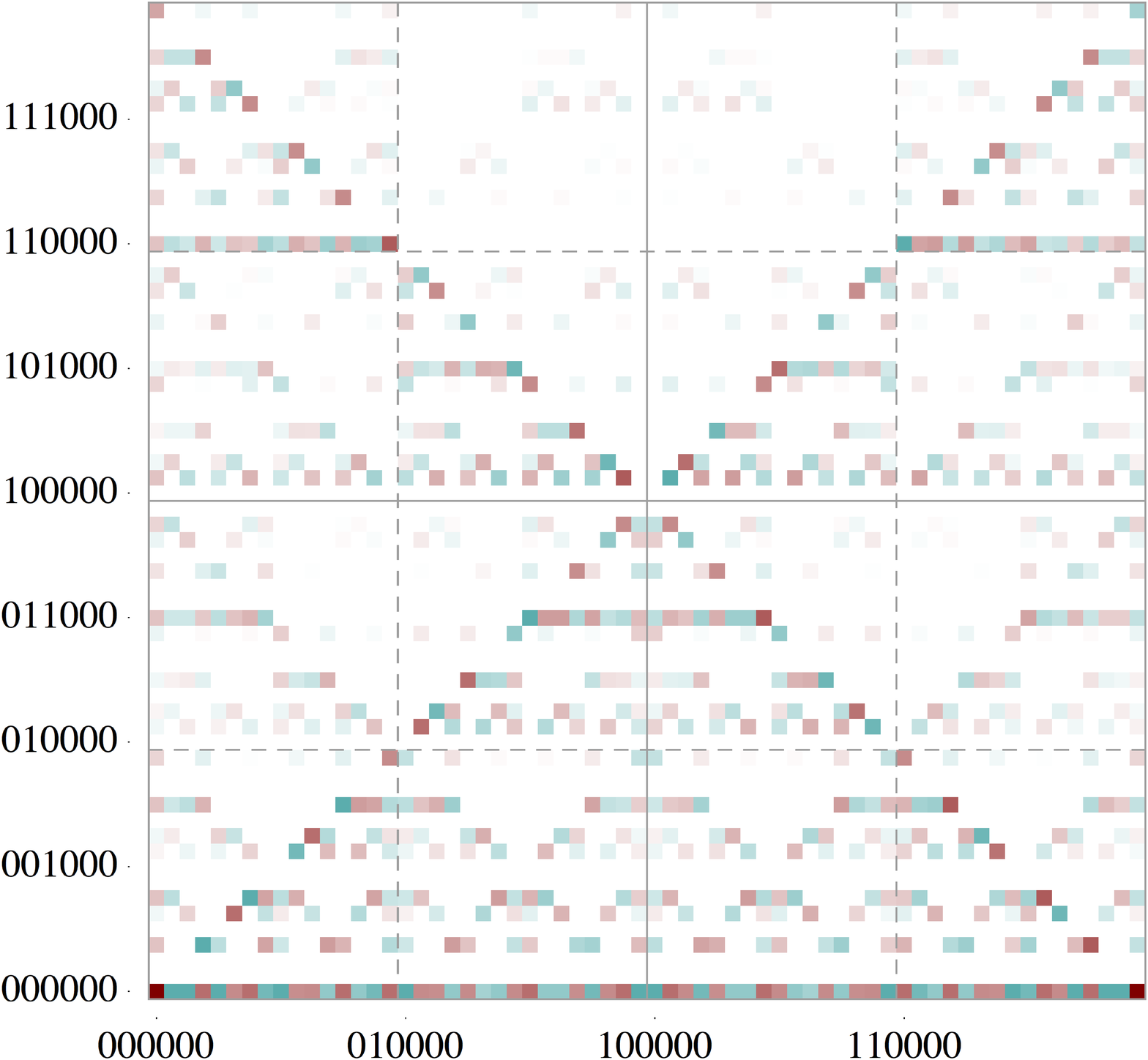,width=5cm} \\
\epsfig{file=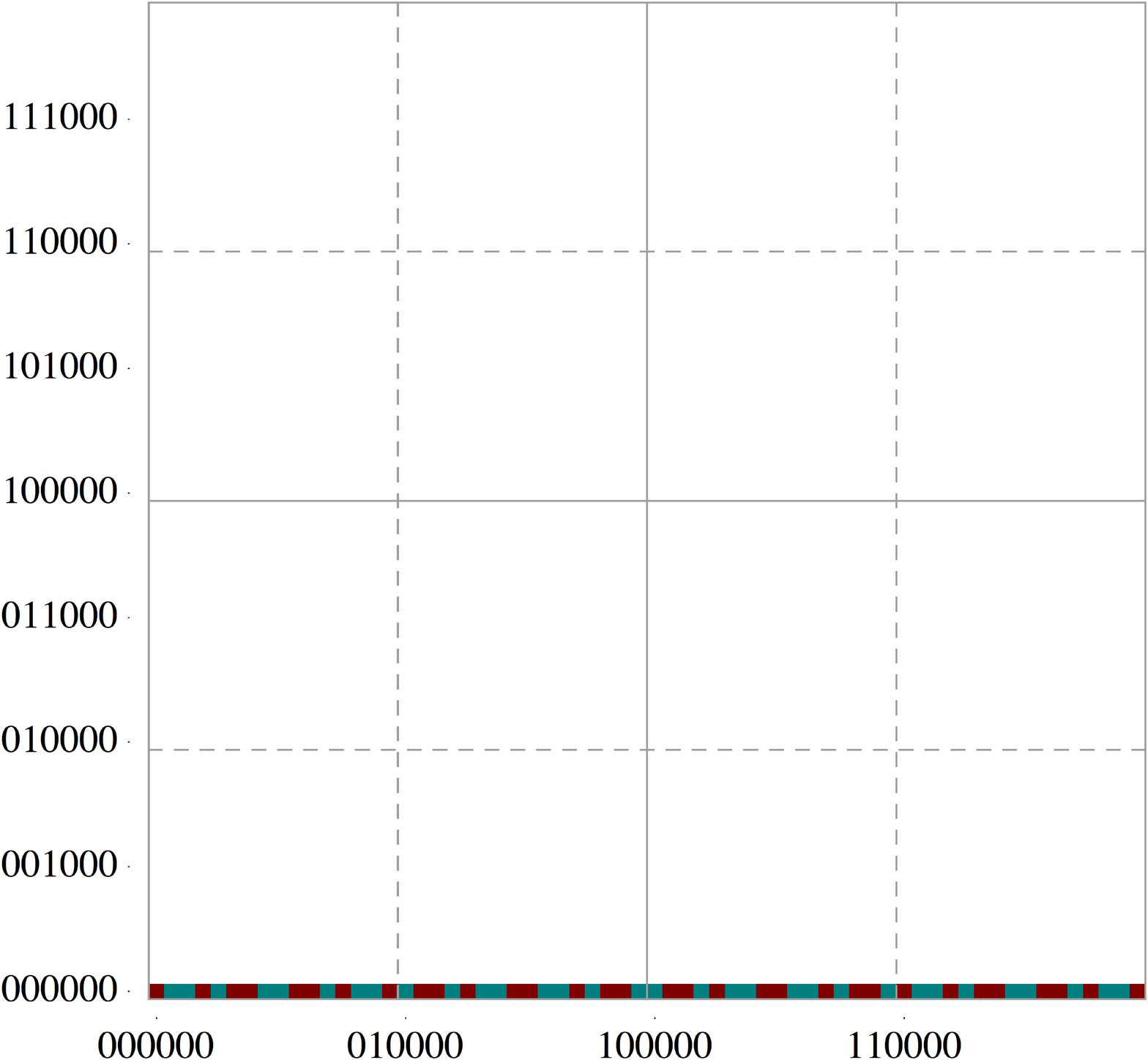,width=5cm}
\end{tabular}
\end{center}
\caption{\label{fig.pauliitf} Frame representation for $N=6$ qubits in
  the ground state of the AF ITF Hamiltonian \ref{itf.model} with
  PBC. The top panel shows the case where $\Gamma\to 0$, and
  correlations are established in the $Z$-axis. The central panel is
  critical, $\Gamma=1$ and the lower one shows the $\Gamma\to\infty$
  case, where correlations are strong in the $X$-axis.}
\end{figure}

As an example, we provide in figure \ref{fig.pauliitf} images
illustrating the ITF quantum phase transition: above, $\Gamma$ is
small and only correlations in the $Z$-axis are relevant. Below,
$\Gamma$ is large and correlations appear only in the $X$-axis. The
middle panel shows the critical case.

\section{Conclusions and further work}

In this work we have described a family of schemes which allow
visualization of the information contained in quantum many-body
wavefunctions, focusing on systems of many qubits. The schemes are
self-similar by design: addition of new qubits results in a higher
resolution of the plots. The thermodynamic limit, therefore,
corresponds to the continuum limit.

The philosophy behind the schemes is to start out with a region $D$
and divide it into several congruent subdomains, all of them similar
to $D$. This subdivision procedure can be iterated as many times as
needed, producing an exponentially large amount of subdomains, each of
them characterized by a geometrical index. This index can be now
associated to an element of the tensor-basis of the Hilbert space, and
its corresponding wavefunction amplitude goes, through a certain color
code, into that subdomain. The most simple example is with $D$ a
square which splits into four equal quadrants, but we can also start
with a right triangle, or even with a line segment.

Physical features of the wavefunctions translate naturally into visual
features of the plot. For example, within the scheme in section
\ref{2dplots}, the spin-liquid character of the ground state of the
Heisenberg model shows itself in a characteristic pattern of diagonal
lines. This pattern is able to distinguish between open and periodic
boundary conditions. Other features which show up in the plots is
magnetization, criticality, invariance under translations or
permutation of the qubits, and Marshall's sign rule. We have analysed
the characteristic features of product states, the ground states of
the Ising model in a transverse field, the Majumdar-Ghosh Hamiltonian
or Dicke states. We have also studied spin-1 systems, such as the AKLT
state.

A very relevant physical feature which becomes apparent in the plots
is entanglement. Factorizability is straightforward to spot: a
wavefunction is factorizable if all sub-images at a certain division
level are equal, modulo normalization. The Schmidt rank of a given
left-right partition of the system is related to the dimension of the
subspace spanned by all sub-images within the corresponding
subdivision of the plot and, so, a crude method to obtain an upper
bound is to count the number of different sub-images. The full
information about entanglement is contained in the matrix that we have
termed as cross-correlation, which contains the overlap between all
subimages at a certain division level.

In a very different spirit, we have illustrated the frame
representations of quantum states of many qubits. This approach is
related to Wooters' group ideas. In it, the expectation values of a
selected set of operators are shown in a 2D array, which is again
displayed in a self-similar manner.

In this work we have taken the first steps in the exploration of an
alternative strategy in the study of quantum many-body sytems, which
can provide support to the corpus of methods in the field. Regarding
further work, we would like to stress the further exploration of
interesting quantum many-body states which we have not done here, for
example the ground states of fermionic Hamiltonians, the Hubbard
model, the Mott transition or the BEC-BCS crossover. Understanding the
plotting structure of matrix product states of low dimension might
also result profitable. Moreover, the mathematical properties of the
mapping itself are worth studying by themselves.

As a final remark, we would like to announce that source code and
further images can be found at {\tt http://qubism.wikidot.com}, a
webpage dedicated to qubism-related resources.


\ack

This work has been supported by the Spanish government grants
FIS2009-11654, FIS2009-12964-C05-01, FIS2008-00784 (TOQATA) and
QUITEMAD, and by ERC grant QUAGATUA. M.L. acknowledges the Alexander
von Humboldt foundation and the Hamburg theory
award. J.R.-L. acknowledges D.~Peralta and S.N.~Santalla for very
useful discussions.

\section*{References}


\end{document}